\documentclass[twocolumn, twocolappendix]{aastex631}
\usepackage{gensymb}
\usepackage{amsmath}

\received{Dec 12, 2022}
\revised{Feb 10, 2023}
\accepted{Feb 27, 2023}

\submitjournal{ApJ}

\graphicspath{{./}{figures/}}

\begin{document}

\title{Probing the cold neutral medium through \ion{H}{1} emission morphology with the scattering transform}

\author[0000-0002-2679-4609]{Minjie Lei}
\affiliation{Department of Physics, Stanford University, Stanford, CA 94305, USA}
\affiliation{Kavli Institute for Particle Astrophysics \& Cosmology, P.O. Box 2450, Stanford University, Stanford, CA 94305, USA}

\author[0000-0002-7633-3376]{S. E. Clark}
\affiliation{Department of Physics, Stanford University, Stanford, CA 94305, USA}
\affiliation{Kavli Institute for Particle Astrophysics \& Cosmology, P.O. Box 2450, Stanford University, Stanford, CA 94305, USA}




\begin{abstract}

Neutral hydrogen (\ion{H}{1}) emission exhibits complex morphology that encodes rich information about the physics of the interstellar medium. We apply the scattering transform (ST) to characterize the \ion{H}{1} emission structure via a set of compact and interpretable coefficients, and find a connection between the \ion{H}{1} emission morphology and \ion{H}{1} cold neutral medium (CNM) phase content. Where \ion{H}{1} absorption measurements are unavailable, the \ion{H}{1} phase structure is typically estimated from the emission via spectral line decomposition. Here, we present a new probe of the CNM content using measures solely derived from \ion{H}{1} emission spatial information. We apply the ST to GALFA-\ion{H}{1} data at high Galactic latitudes ($\left|b\right|>30\degree$), and compare the resulting coefficients to CNM fraction measurements derived from archival \ion{H}{1} emission and absorption spectra. We quantify the correlation between the ST coefficients and measured CNM fraction ($f_{\mathrm{CNM}}$), finding that the \ion{H}{1} emission morphology encodes substantial $f_{\mathrm{CNM}}$-correlating information and that ST-based metrics for small-scale linearity are particularly predictive of $f_{\mathrm{CNM}}$. This is further corroborated by the enhancement of the $I_{857}/N_{H\,I}$ ratio with larger ST measures of small-scale linearity. These results are consistent with the picture of regions with higher CNM content being more populated by small-scale filamentary \ion{H}{1} structures. Our work illustrates a physical connection between the \ion{H}{1} morphology and phase content, and suggests that future phase decomposition methods can be improved by making use of both \ion{H}{1} spectral and spatial information. 

\end{abstract}



\newpage

\section{Introduction} \label{sec:intro}

The interstellar medium (ISM) is a diffuse multiphase structure that fills most of the Milky Way volume. The ISM gas is generally made up of ionized, molecular, and atomic components that are linked in a dynamic interplay that is responsible for important Galactic processes across scales. The atomic component, composed primarily of neutral hydrogen (\ion{H}{1}), plays a crucial role in the life cycles of galaxies. \ion{H}{1} is the progenitor material from which dense star-forming molecular clouds form \citep{clarkp12-mc, inoue12-mc, steinberg14-mc, lee15-mc}. The neutral medium is further composed of a cold neutral medium (CNM), warm neutral medium (WNM), and a thermally unstable medium \citep{field69-ph, wolfire03-hi, kalberla09-hi}. Understanding the \ion{H}{1} phase distribution of the Milky Way, and how matter and energy are transferred between phases, are fundamental questions in ISM research. 

The primary observational probe of the interstellar gas distribution is the 21cm line from the hyperfine transition of ground-state neutral hydrogen. Mapping the Galactic neutral gas distribution with the 21cm line has enabled fruitful analyses of Milky Way's structure and dynamics \citep{kalberla09-hi}. However, an accurate determination of \ion{H}{1} properties, like temperature and density, involves the unique challenge of requiring both emission and absorption measurements \citep{Heiles2003-ca}. Observational studies reveal that the 21cm emission can be mapped along every line of sight (LOS) \citep{kalberla09-hi, winkel16-hi}. However, absorption observations are limited by the availability of background radio continuum sources. The currently-available absorption measurements \citep{Heiles2003-ca, Murray2015-yu, Murray2018-xe} are too sparse to resolve the spatial distribution of \ion{H}{1} phases over most of the sky \citep[except in special circumstances, e.g.,][]{mg06-cg}. Future surveys with the Square Kilometer Array \citep[SKA;][]{McClure-Griffiths:2015}, and ongoing observations like the Galactic Australian SKA Pathfinder (GASKAP) Survey \citep{dickey13-gs}, will significantly increase the number of available sightlines with absorption measurements. However, the availability of background sources still fundamentally limits our ability to spatially resolve \ion{H}{1} phase information. 

The lack of absorption measurements for directly determining of \ion{H}{1} properties has prompted the development of methods for estimating the \ion{H}{1} properties using 21cm emission data alone. A common approach is Gaussian phase decomposition \citep{matthews57-on, takakubo66-nh, mebold72-ot, haud07-gd, kalberla18-po}, where emission spectra are decomposed into Gaussian components, with the amplitude and linewidth of each Gaussian then being used to infer the physical properties of each feature. However, in general, there is no unique solution to the decomposition, with further complications coming from systematics, velocity blending, and non-Gaussian line shapes. To address these problems, recent works have improved Gaussian decomposition, using innovative approaches like regularization and automated component selection \citep{marchal19-rs, riener20-ag}. Enabled by increasingly realistic simulations of the ISM \citep{Kim2014-db}, methods using deep learning approaches, like convolutional neural networks (CNNs) have also been developed, and perform comparably to the state-of-the-art Gaussian decomposition methods \citep{Murray2020-uf}. 

Existing phase decomposition methods mostly assume that \ion{H}{1} phases can be inferred from the velocity structure of the 21cm emission. Some algorithms do make use of limited spatial information in the immediate region surrounding each sightline, e.g. to ensure the spatial coherence of the derived parameters \citep{marchal19-rs} or to estimate the expected emission spectra in the presence of a continuum source \citep{Murray2020-uf}. However, these methods do not directly use the spatial morphology of the \ion{H}{1} emission to inform the phase decomposition.  In recent years, with the increasing availability of \ion{H}{1} emission observations, including some large-scale high-spatial-resolution observational surveys, there has been growing interest in deriving the properties of the ISM from the \ion{H}{1} spatial morphology, with fruitful results. The physics probed by the \ion{H}{1} morphology includes $\mathrm{H_2}$ formation \citep{barriault10-hm}, galaxy dynamics \citep{soler20-gd, soler22-hm}, high-velocity cloud instabilities \citep{barger20-hm}, galactic outflows \citep{mg18-hm,teodoro20-of}, star formation \citep{hacar22-ic, yu22-hm}, and the structure of the interstellar magnetic field, with applications to cosmological foregrounds \citep{clark14-ma, clark15-fm, kalberla16-hm, clark18-np, clarkh19-mp, halal22-hi}. In particular, the study of \ion{H}{1} intensity maps using machine vision algorithms, like the Rolling Hough Transform \citep[RHT;][]{clark14-ma}, has revealed remarkably linear filamentary features that align with magnetic field directions, as traced by dust and starlight polarization \citep{clark14-ma, clark15-fm, kalberla16-hm, clarkh19-mp}. These small-scale filamentary structures have been further shown to be preferentially associated with the CNM \citep{kalberla18-po, clark19_pn, peak19-cl, Murray2020-uf}. Similar cold magnetically aligned structures have also been identified in absorption \citep{mg06-cg}. These results suggest that substantial \ion{H}{1} phase-correlating information could also be encoded in the spatial structure of the 21cm emission, in addition to the spectral structure. Thus, an effective statistical description of the \ion{H}{1} emission morphology, combined with existing methods of extracting phase information from the spectral dimension, could potentially result in significantly improved phase separation performance.

In this work, we apply a flexible quantification of the HI emission morphology, and establish that the spatial structure of the HI emission is quantifiably predictive of the CNM fraction using a general morphological framework. We characterize the \ion{H}{1} morphology by using the scattering transform (ST), which is a powerful statistical technique capable of extracting significant non-Gaussian information into a set of compact and interpretable coefficients. Recent applications of the ST include ISM studies of non-Gaussian structures in dust emission \citep[e.g.][]{Robitaille2014-tj, allys2019-tr, blanchard20-st, saydjari21-st, delouis22-st} as well as cosmological parameter inference in contexts such as large-scale structure \citep{cheng20-an, valo22-gb} and line intensity mapping \citep{chung2022-eo, greig22-st}. Compared to dedicated filament finders like the RHT, the ST is a far more flexible and general descriptor of field morphology. Here, we explore the first application of the ST to deriving a set of morphological measures from the \ion{H}{1} emission structure, and report correlations with the CNM fraction ($f_{\mathrm{CNM}}$) derived from archival emission and absorption measurements.

The rest of the paper is organized as follows. In Section \ref{sec:data}, we detail the \ion{H}{1} emission and absorption data used in the study. In Section \ref{sec:st}, we introduce the ST. In Section \ref{sec:hi_morph_res} we then discuss the results of applying the ST to the \ion{H}{1} maps and the physical interpretation of these results. The main results of the connection between the \ion{H}{1} morphology and $f_{\mathrm{CNM}}$ are presented in Section \ref{sec:fcnm_corr_res}, followed by a discussion and conclusions in Sections \ref{sec:discussion} and \ref{sec:conclusion}. Further technical discussions can be found in Appendices \ref{appx:systematics} and \ref{appx:st_synthesis}.

\section{Data} \label{sec:data}

\begin{figure*}[t]
\centering
\includegraphics[width=0.8\textwidth]{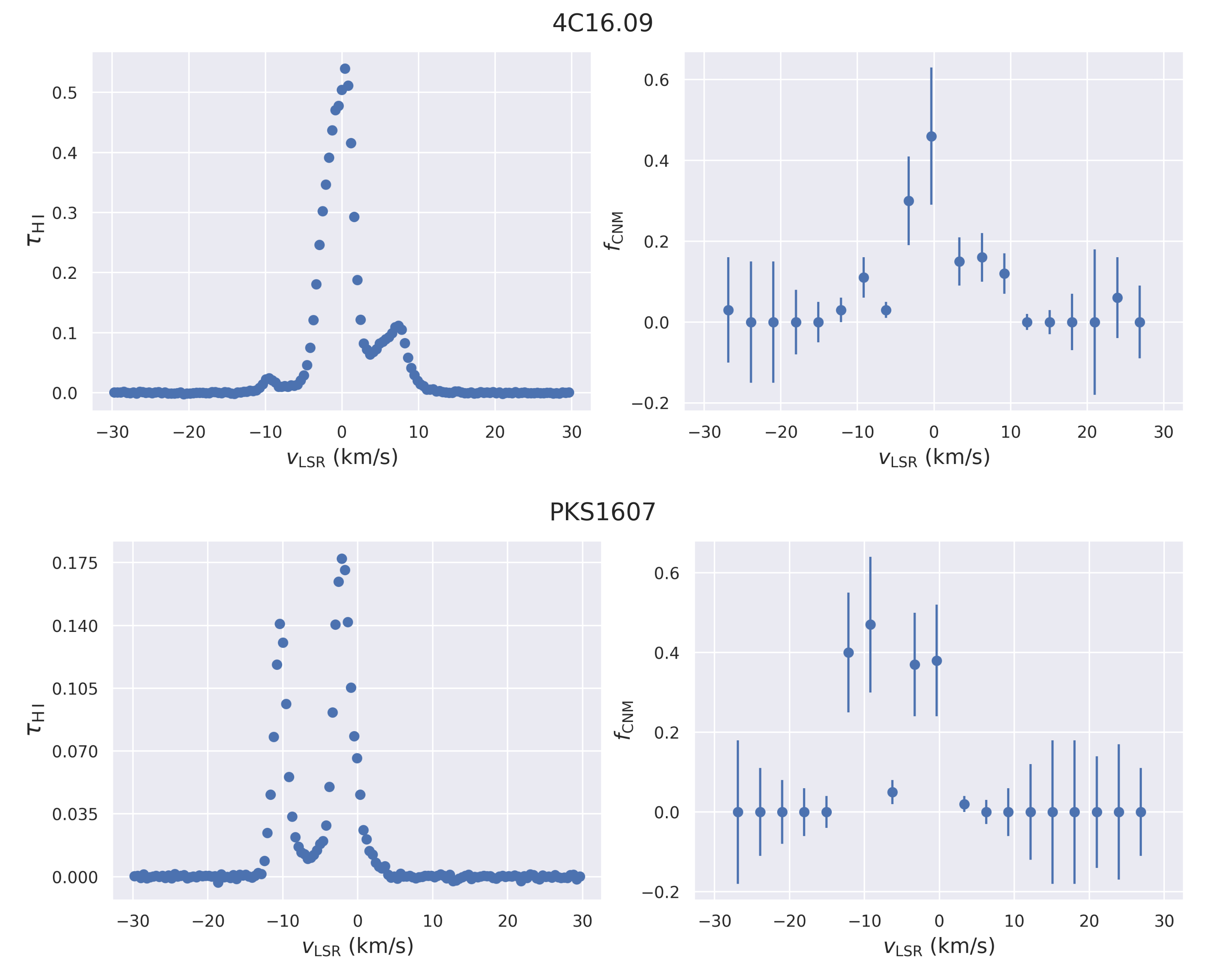}
\caption{$f_{\text{CNM}}(v)$ spectra (right), with channel width $v=3$ km/s computed from the spectral line pairs ($\tau(v)$, $T_b(v)$) as described in Section \ref{subsec:fcnm_data}, with corresponding optical depth spectra $\tau(v)$ (left), for two example sightlines, 4C16.09 and PKS1607.}
\label{fig:fcnm_spec}
\end{figure*}

\begin{figure*}[t]
\centering
\includegraphics[width=0.9\textwidth]{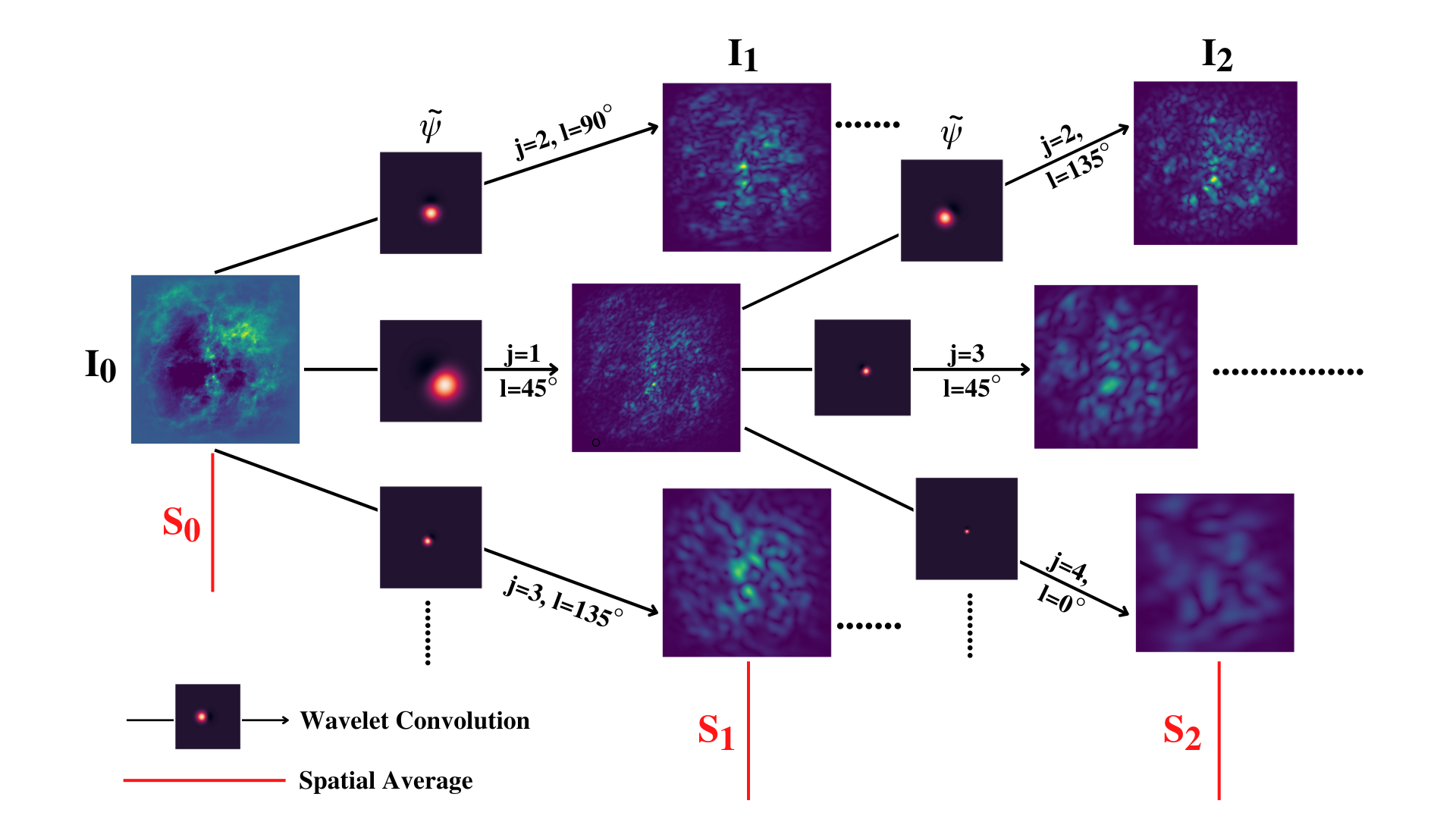}
\caption{Illustration of applying the ST process to a sample GALFA-\ion{H}{1} column density patch. The sample image is centered at $(l=220\degree, b=20\degree)$, spanning $34\degree$ on each side. The ST results are shown for up to two orders. At each order, we explicitly plot the intermediate images resulting from the convolution with the wavelets $\psi(j, l)$ for selected scales and orientations, with the corresponding wavelets being shown in Fourier space. The detailed formalism and number of coefficients are discussed in Section \ref{subsec:st_form}.}
\label{fig:st_flow}
\end{figure*}

\subsection{\ion{H}{1} Emission} \label{subsec:hi_ems}
The \ion{H}{1} emission dataset used in our analysis is the Data Release 2 of the Galactic Arecibo $L$-band Feed Array Survey \citep[GALFA-\ion{H}{1};][]{Peek2017-ql}. The GALFA-\ion{H}{1} survey covers $\sim32\%$ of the sky, from decl. $-1\degree17\arcmin$ to decl. $+37\degree57\arcmin$ across all R.A. It has the highest angular (4\arcmin) and spectral (0.184 km/s) resolution of any large-area Galactic 21 cm emission survey to date. We combine the high-resolution emission spectra with \ion{H}{1} absorption measurements into spectral line pairs, to derive the LOS $f_{\text{CNM}}$ data used in our analysis. Moreover, the high angular resolution enables a detailed ST technique that is described in more detail in the following sections. 

\subsection{\ion{H}{1} Absorption} \label{subsec:hi_abs}
For the 21cm absorption measurements, we adopt the same set of samples described in \citet{Murray2020-uf}, including 58 high-latitude sightlines assembled from the Spectral Line Observations
of Neutral Gas with the Karl G. Jansky Very Large Array survey \citep[21-SPONGE;][]{Murray2018-xe} and the Millennium Survey with Arecibo \citep{Heiles2003-ca}. From 21-SPONGE, 30 spectra in the high Galactic latitude ($|b|>30\degree$) GALFA-\ion{H}{1} sky ($-1\degree17\arcmin$ $<$ decl. $<$ $+37\degree57\arcmin$, 0 $<$ R.A $<$ $360\degree$) are selected, with high optical depth sensitivity ($\sigma_{\tau\mathrm{H\,I}}$) and velocity resolution ($\delta v$ = 0.42 km/s). The additional 28 spectra are selected from the Millennium Survey to be unique relative to the 21-SPONGE sightlines in the Arecibo footprint, and do not have significant systematics or spectral artifacts.

\subsection{CNM Fraction} \label{subsec:fcnm_data}
From the 21cm emission and absorption data set we construct spectral line pairs to derive the \ion{H}{1} properties, including the mass fraction of the CNM ($f_{\mathrm{CNM}}$). We follow the procedure of spectral line pair construction described in \citet[][Section 2.3]{Murray2020-uf}, where the \ion{H}{1} absorption $\tau_{\mathrm{H\,I}}(v)$ spectra and \ion{H}{1} emission GALFA-\ion{H}{1} cubes are smoothed to the same 0.42 km/s channel resolution of 21-SPONGE. To estimate the expected emission spectra in the absence of each background continuum source, we extract brightness temperature spectra ($T_B(v)$) from a $9\arcmin\times9\arcmin$ patch around each source, subtracting the inner $3\arcmin\times3\arcmin$ region containing the source. The channel-dependent uncertainties $\sigma_{\tau_{\mathrm{HI}}}(v)$ and $\sigma_{T_B}(v)$ are estimated using methods described in \citet{Murray2015-yu}. Motivated by their result, of high-velocity structures not containing significant enough absorption to affect the estimated $f_{\mathrm{CNM}}$, we restrict the spectral pairs to velocities $|v_{\mathrm{LSR}}| < 40$ km/s. This results in 58 spectral line pairs $(\tau_{\mathrm{H\,I}}(v), T_B(v))$ and their uncertainties $(\sigma_{\tau_{\mathrm{HI}}}(v), \sigma_{T_B}(v))$. 

To compute $f_{\mathrm{CNM}}$ from the spectral pairs and their uncertainties, we again follow the steps outlined in \citet{Murray2020-uf}: 
\begin{equation} \label{eq:fcnm}
    f_\mathrm{CNM} \approx \frac{T_{\mathrm{CNM}}}{\left<T_s\right>} \frac{T_{s, \mathrm{WNM}}-\left<T_s\right>}{T_{s,\mathrm{WNM}}-T_\mathrm{CNM}},
\end{equation}
where $T_{\mathrm{CNM}}$ is the CNM kinetic temperature, $T_{s, \mathrm{WNM}}$ is the reference WNM spin temperature, and $\left<T_s\right>$ is the optical depth-weighted average spin temperature along the LOS. Following \citet{Murray2020-uf}, we choose $T_{\mathrm{CNM}}=50\mathrm{K}$ and $T_{s, \mathrm{WNM}}=1500\mathrm{K}$, which have been shown to be good approximations for the local ISM \citep{Kim2014-db, Murray2018-xe}. Here, $T_{s, \mathrm{WNM}}$ is not chosen to correspond to the real spin temperature of the WNM, but as a reference temperature above which $f_{\mathrm{CNM}}$ is zero. $\left<T_s\right>$ is given by 
\begin{equation} \label{eq:ts_avg}
    \left<T_s\right> = \frac{\int \tau_{\mathrm{H\,I}}\ T_s\ dv}{\int \tau_{\mathrm{H\,I}}\ dv}.
\end{equation}
In the isothermal approximation, the spin temperature at a given velocity channel can be computed from the spectral pairs $(\tau_{\mathrm{H\,I}}(v), T_B(v))$ as 
\begin{equation} \label{eq:ts_iso}
    T_s = \frac{T_B(v)}{1-e^{-\tau_{\mathrm{H\,I}}(v)}}.
\end{equation}
We determine the uncertainty in $f_{\mathrm{CNM}}$ by using a simple Monte Carlo error propagation procedure from $(\sigma_{\tau_{\mathrm{HI}}}(v), \sigma_{T_B}(v))$. For our analysis, we compute $(f_{\mathrm{CNM}}, \sigma_{f_{\mathrm{CNM}}})$ along a given LOS, by integrating $\left<T_s\right>$ in Equation \ref{eq:ts_avg} over the full velocity range $|v_{\mathrm{LSR}}| < 40\ \mathrm{km/s}$, as well as $(f_{\mathrm{CNM}}(v), \sigma_{f_{\mathrm{CNM}}}(v))$ by integrating $\left<T_s\right>$ over narrow velocity ranges with channel widths of either $\delta v = 1.5~\mathrm{km/s}$ or $3.0\ \mathrm{km/s}$. Our velocity-integrated $f_{\mathrm{CNM}}$ results are consistent with \citet{Murray2020-uf}. The computed $f_{\mathrm{CNM}}(v)$ spectra are shown with their corresponding $\tau_{\mathrm{H\,I}}(v)$ spectra in Figure \ref{fig:fcnm_spec} for two example sightlines. In the following sections, we will explore the correlation of morphology measures derived using the ST with both the LOS-integrated $f_{\mathrm{CNM}}$ data and the narrow velocity channel $f_{\mathrm{CNM}}(v)$ spectra. 

\section{ST Formalism} \label{sec:st}
In this section, we summarize the formalism of the ST, the motivation for using it as a morphological measure, and the intuitive interpretations of its coefficients. 

\subsection{Motivation and Formulation} \label{subsec:st_form}
The ST was originally proposed by \citet{mallat2012-gi} in the context of signal processing in computer vision, to extract information from high-dimensional input fields. There has been growing interest in applying the ST to astrophysical data analysis \citep[e.g.][]{Robitaille2014-tj, allys2019-tr, cheng20-an, blanchard20-st, chung2022-eo, delouis22-st, valo22-gb, greig22-st}. This derives from the ST's capacity to encode substantial non-Gaussian information, in a set of coefficients that are intuitively meaningful. Compared with other approaches to capturing non-Gaussian information using higher-order statistics, such as the N-point function, the ST is more robust to small perturbations and geometric deformations, and captures a more compact set of descriptors relative to image size \citep{cheng2021-si}. Moreover, the mathematical formulation of the ST shares similarities with CNNs. The analysis of wavelet scattering has provided key insights into the properties of CNNs, specifically how the network coefficients relate to image sparsity and geometry \citep{bruna2012-jo}. 

In the ST formulation, to extract information from an input field $I(\boldsymbol{x})$, a scattering operation composed of a wavelet convolution followed by a modulus step is applied iteratively, to generate a group of output fields. The scattering coefficients are then defined as the expectation values of these fields, and together these coefficients characterize the statistical properties of the original field. We illustrate the process of applying the ST to a sample field in Figure \ref{fig:st_flow}. Formally, under one iteration of the scattering step, the input field $I(\boldsymbol{x})$ will be transformed as 
\begin{equation} \label{eq:scatter_step}
    I_1^{j_1, l_1}(\boldsymbol{x}) = \left|I(\boldsymbol{x})*\psi_{j_1, l_1}(\boldsymbol{x})\right|
\end{equation}
where $*$ denotes convolution and $\psi_{j_1, l_1}(\boldsymbol{x})$ is a localized oriented wavelet probing scale $j_1$ and orientation $l_1$. We follow \citet{cheng2021-si} and use the Morlet wavelet for our study \citep[Appendix B]{cheng2021-si}. Taking the expectation value of $I_1(\boldsymbol{x})$ yields the first-order scattering coefficients $S_1(j_1, l_1)$. Combined with a family of wavelets $\psi_{j, l}(\boldsymbol{x})$, the successive application of the scattering operation produces a hierarchy of coefficients that probe scale and orientation interactions at increasing order:
\begin{equation} \label{eq:scatter_step}
    I_n^{j_1,...,j_n,l_1,...,l_n}(\boldsymbol{x}) = \left|I_{n-1}(\boldsymbol{x}) * \psi_{j_n, l_n}(\boldsymbol{x})\right|
\end{equation}
Even though this allows us to capture the ever-growing clustering and complexity, in practice, most physical fields fall in the regime where most of the variance is stored in the lower-order scattering coefficients \citep[Section 4.1]{cheng2021-si}. Thus, in this analysis, we will only work with scattering coefficients up to second-order $S_2$, which are given by:
\begin{eqnarray} \label{eq:st_n2}
    S_0 =& \left<\left|I(\boldsymbol{x})\right|\right> \nonumber \\ 
    S_1(j_1, l_1) =& \left<\left|I(\boldsymbol{x})*\psi_{j_1, l_1}(\boldsymbol{x})\right|\right> \\ \nonumber
    S_2(j_1, j_2, l_1, l_2) =& \left<\left|(\left|I(\boldsymbol{x})*\psi_{j_1, l_1}(\boldsymbol{x})\right|) * \psi_{j_2, l_2}(\boldsymbol{x})\right|\right> 
\end{eqnarray} 
where the choice of $j$ ranges over a dyadic sequence of scales $2^j$, with $1\leq j\leq J$. The maximum scale $J$ is less than $\log_2{N}$ where $N$ is the dimension of the input field. Furthermore, only combinations with $j_2 > j_1$ carry significant physical information. For the $j_2 \leq j_1$ coefficients, the second scattering operation acts as a bandpass filter on the scales that have already been suppressed by the first scattering, resulting in information loss \citep[Section 3.1]{cheng2021-si}. In the 2D input field case, the orientation parameter $l$ corresponds to wavelets with angular sizes $\pi/L$ and position angles $\pi l/L$, over the range $0 \leq l < L$. Thus, for the $n=2$ example of $S_2(j_1, j_2, l_1, l_2)$ coefficients with maximum scale and orientation $J$ and $L$, we have a total of $J(J-1)L^2/2$ coefficients. The logarithmic sampling of the scales and the discrete orientation selection ensure that the number of scattering coefficients grows slowly with the field size, resulting in a dense set of descriptors. In practice, further reductions are often applied to the full scattering coefficients, to arrive at a set of more readily interpretable coefficients that are tailored for specific applications. In the following section, we describe the scattering coefficients and their interpretation as adapted to our task of probing the CNM content. 

\begin{figure}[t]
\centering
\includegraphics[scale=0.22]{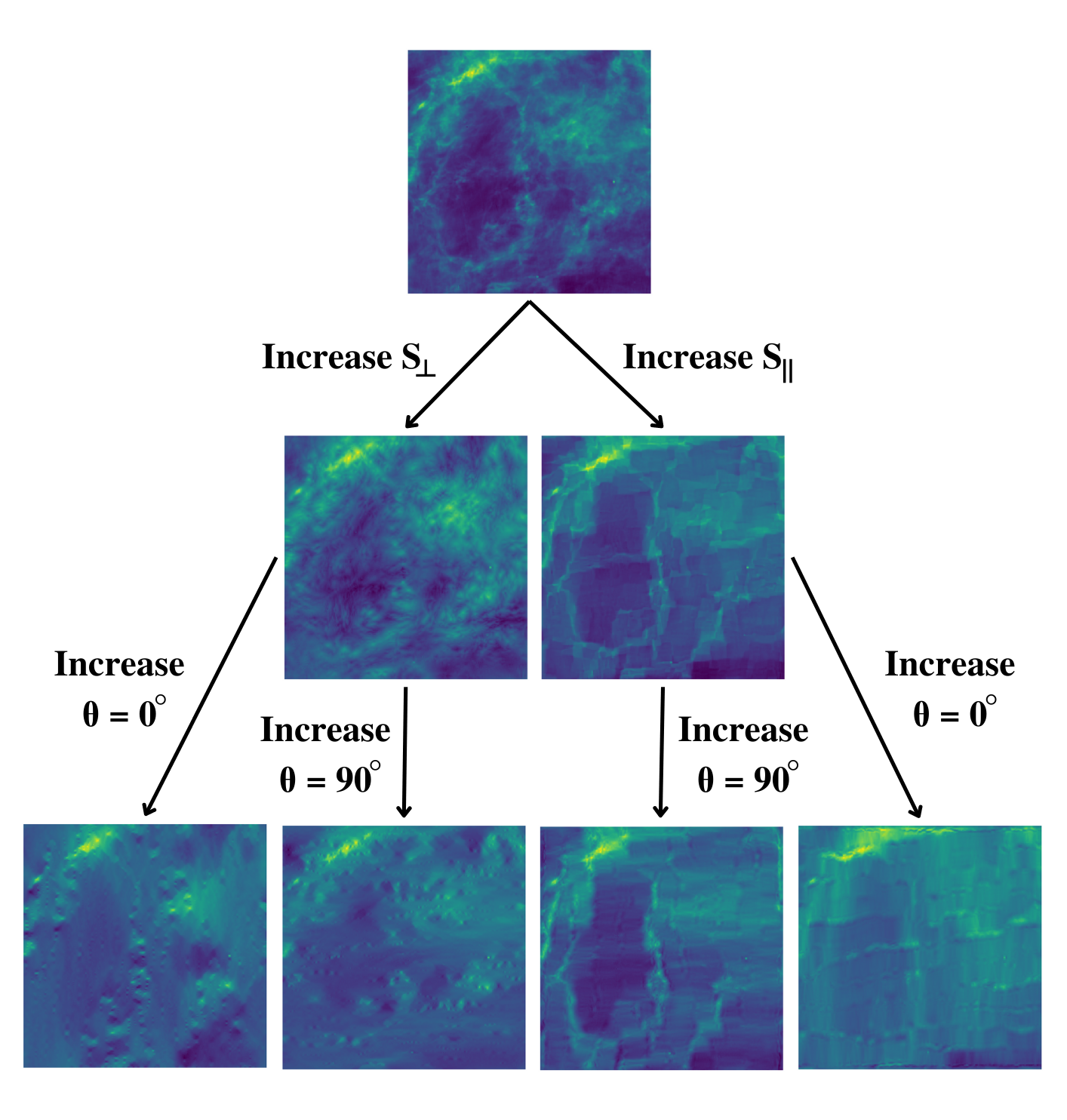}
\caption{Synthesized images seeded from the same GALFA-\ion{H}{1} sample region, constructed by varying the ST coefficients of the original image via gradient descent. The same GALFA-\ion{H}{1} image is used as in Figure \ref{fig:st_flow}, resampled to $64\times64$ pixels. From the seed image, increasing the ST coefficients with the alignment orientation $\Delta \theta = 90\degree$ (higher $S_\perp$) generates softer features, while increasing $\Delta \theta = 0\degree$ (higher $S_\parallel$) generates more linear textures. At the second level, these textures are enhanced at an absolute orientation of either $\theta = 0\degree$ or $90\degree$.}
\label{fig:st_syn}
\end{figure}

\subsection{Reduction and Interpretation} \label{subsec:st_reduc}
One of the key motivations for adopting the ST to characterize fields is its interpretability. To start, the zeroth-order scattering coefficient $S_0$ is just the mean of the field $\left<\left|I(\boldsymbol{x})\right|\right>$. The first-order coefficients $S_1(j_1, l_1)$, which result from one iteration of a wavelet convolution followed by a modulus operation, are qualitatively similar to the power spectrum. Both characterize field fluctuations as a function of scale, with the difference being that the ST employs convolution with a family of localized wavelets and uses the $L^1$ norm, instead of the $L_2$ norm, of the convolved fields. This has the benefit of not amplifying fluctuations as much and minimizing the variance of the estimator: properties that contribute to the ST being more robust to noise and outliers than higher-order statistics. The second-order coefficients $S_2(j_1, j_2, l_1, l_2)$, which are the result of two successive scattering operations, capture the scale and orientation interactions of the field and carry substantial non-Gaussian information. These coefficients quantify the strength of fluctuations mapped in the first-order output $I_1^{j_1, l_1}(\boldsymbol{x})$. Thus, intuitively, the coefficients characterize how features at a smaller scale $j_1$ along the orientation $l_1$ cluster on a larger-scale $j_2$ along orientation $l_2$. 

The full set of $S_0$, $S_1$, $S_2$ coefficients contains a total of $1+JL+J(J-1)L^2/2$ parameters, which are reasonably compact, but in specific applications the coefficients will be highly correlated and can be further reduced into a more efficient set of summary statistics. Motivated by the discussion in \citet[Section 4]{cheng2021-si}, we apply the following normalization and series of reductions to the full set of parameters. First, since each $S_2$ coefficient is proportional to the corresponding $S_1$ coefficients by construction, we decorrelate $S_2$ by applying the normalization $\tilde{S}_2 \equiv S_2 / S_1$. Henceforth, we will drop the tilde, and use $S_2$ to denote the decorrelated coefficients. 
We then reorder the parameters as follows: 
\begin{equation} \label{eq:st_normal_order}
    S_2(j_1, j_2, \Delta\theta, \theta) = S_2(j_1, j_2, |l_1-l_2|, l_1)
\end{equation}
where $\Delta\theta = |l_1-l_2|$ has the more intuitive interpretation of being the alignment angle between the orientation of features on scale $j_1$ and the orientation of their clustering on the larger-scale $j_2$. For example, large $\Delta\theta = 0\degree$ components correspond to fluctuations at one scale, being distributing along the same direction on a larger scale, which is the case for linear/filamentary features. $\theta = \l_1$ then denotes the absolute orientation of features with a certain alignment, e.g., large $(\Delta\theta=0\degree, \theta=90\degree)$ components correspond to predominantly linear features along the vertical direction. From the decorrelated, reordered coefficients, we consider the following series of reduced coefficients:
\begin{itemize}
    \item Full $S_2$ coefficients [$J(J-1)L^2/2$ parameters]: 
    \begin{eqnarray} \label{eq:st_full}
        S_2(j_1, j_2, \Delta\theta, \theta)
    \end{eqnarray}
    \item $S_2^{\mathrm{iso}}$ coefficients [$J(J-1)L/2$]: 
    \begin{eqnarray} \label{eq:st_full_iso}
        S_2(j_1, j_2, \Delta\theta) = \left<S_2(j_1, j_2, \Delta\theta, \theta)\right>_{\theta}
    \end{eqnarray}
    \item $S_{2\parallel}, S_{2\perp}$ coefficients [$2\,*\,J(J-1)L/2$]: 
    \begin{eqnarray} \label{eq:st_po}
        S_{2\parallel}(j_1, j_2, \theta) = S_2(j_1, j_2, \Delta\theta=0\degree, \theta)\\
        S_{2\perp}(j_1, j_2, \theta) = S_2(j_1, j_2, \Delta\theta=90\degree, \theta)
    \end{eqnarray} 
    \item $S_{2\parallel}^{\mathrm{iso}}$ and $S_{2\perp}^{\mathrm{iso}}$ coefficients [$2\,*\,J(J-1)/2$]:
    \begin{eqnarray} \label{eq:st_po_iso}
        S_{2\parallel}^{\mathrm{iso}}(j_1, j_2) =  \left<S_{2\parallel}(j_1, j_2, \theta)\right>_{\theta}\\
        S_{2\perp}^{\mathrm{iso}}(j_1, j_2) =\left<S_{2\perp}(j_1, j_2, \theta)\right>_{\theta}
    \end{eqnarray}
\end{itemize}
To illustrate the interpretations of these coefficients, we produce $64\times64$ pixel synthesized images with different values of the reduced ST coefficients, as shown in Figure \ref{fig:st_syn}. Starting from a sample GALFA-\ion{H}{1} patch as a seed, the field is gradually varied through gradient descent, to minimize the difference between the ST coefficients of the synthesized image and the provided coefficients. More details about the synthesized fields are provided in Appendix \ref{appx:st_synthesis}. In Figure \ref{fig:st_syn}, to first demonstrate the interpretation of the relative orientation $\Delta \theta$, we vary the $S_{2\parallel}^{\mathrm{iso}}$ and $S_{2\perp}^{\mathrm{iso}}$ coefficients, where the absolute orientation $\theta$ has been averaged over. The results show that the synthesized image with large $S_{2\parallel}^{\mathrm{iso}}$ components has a more linear/filamentary texture than the seed image, while the image with large $S_{2\perp}^{\mathrm{iso}}$ components appears more soft and diffuse. This aligns with our expectation that $S_{2\parallel}^{\mathrm{iso}}$ corresponds to features at scale $j_1$ that cluster along parallel directions to form linear features, while large $S_{2\perp}^{\mathrm{iso}}$ results in features aligning orthogonally along large scales, forming blobs and softer textures. Then to illustrate the absolute orientation $\theta$, the images at the next level down are varied to have large $\theta=0\degree$ vs. $\theta=90\degree$ components, respectively. For $S_{2\parallel}^{\mathrm{iso}}$ coefficients, this corresponds to images with linear features along the horizontal ($\theta=0\degree$) vs. vertical ($\theta=90\degree$) directions, respectively. This correspondence is reversed for the $S_{2\perp}^{\mathrm{iso}}$ coefficients, since $S_{2\perp}^{\mathrm{iso}}(\theta=0\degree)$ describes horizontally oriented small-scale features aligning perpendicular to their original direction at large scales, to form soft features along the vertical direction. Similarly, $S_{2\perp}^{\mathrm{iso}}(\theta=90\degree)$ describes soft features along horizontal directions. In the following sections, we compute this series of ST coefficients on GALFA-\ion{H}{1} data, and examine what they tell us about the morphology of the \ion{H}{1} emission and its correlation with the CNM content. 

\begin{figure*}[t]
    \centering
    \begin{minipage}{\linewidth}
        \centering
        \includegraphics[width=0.83\textwidth]{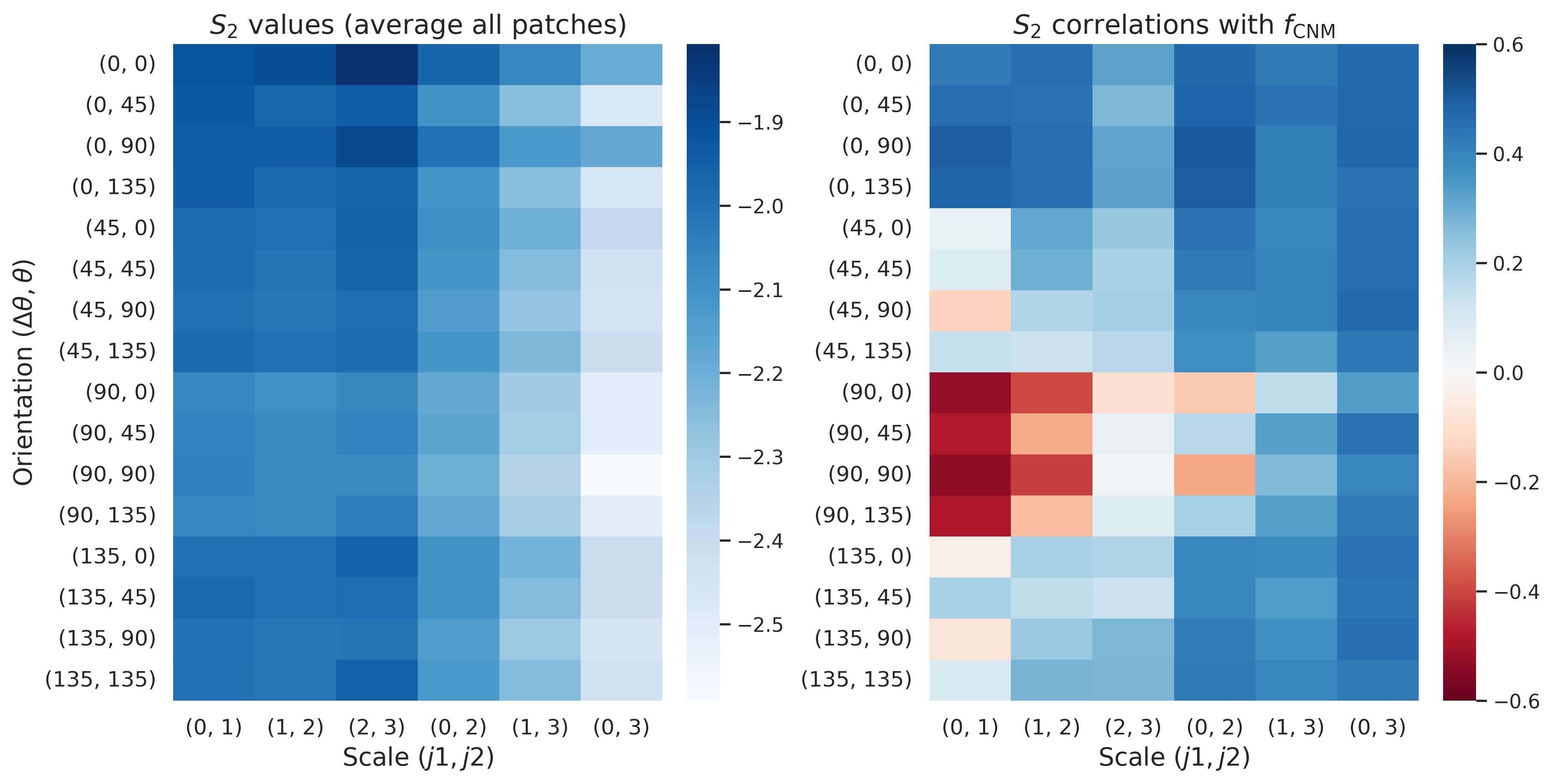}
        \caption{Left: the full $S_2(j_1, j_2, \Delta\theta, \theta)$ coefficients averaged over all GALFA-\ion{H}{1} patches, shown in log scale. Along the scale dimension, higher average $S_2$ values are concentrated at smaller scales. Along the orientation dimension, averaged parallel $\Delta \theta=0\degree$ components are slightly larger than the perpendicular $\Delta\theta=90\degree$ coefficients. Right: correlation (Spearman rank order) of these coefficients with $f_{\mathrm{CNM}}(v)$ for the sightline at the center of each patch. The x-axis varies over the scale components $(j_1, j_2)$, ordered by $|j_2-j_1|$, while the y-axis varies over the orientation components $(\Delta\theta, \theta)$. Small-scale parallel and perpendicular components are shown to be strongly correlated and anticorrelated with $f_{\mathrm{CNM}}$, respectively, in the correlation plot on the right.} \label{fig:st_val_corr}
    \end{minipage}
    \begin{minipage}{\linewidth}
        \centering
        \includegraphics[width=0.85\textwidth]{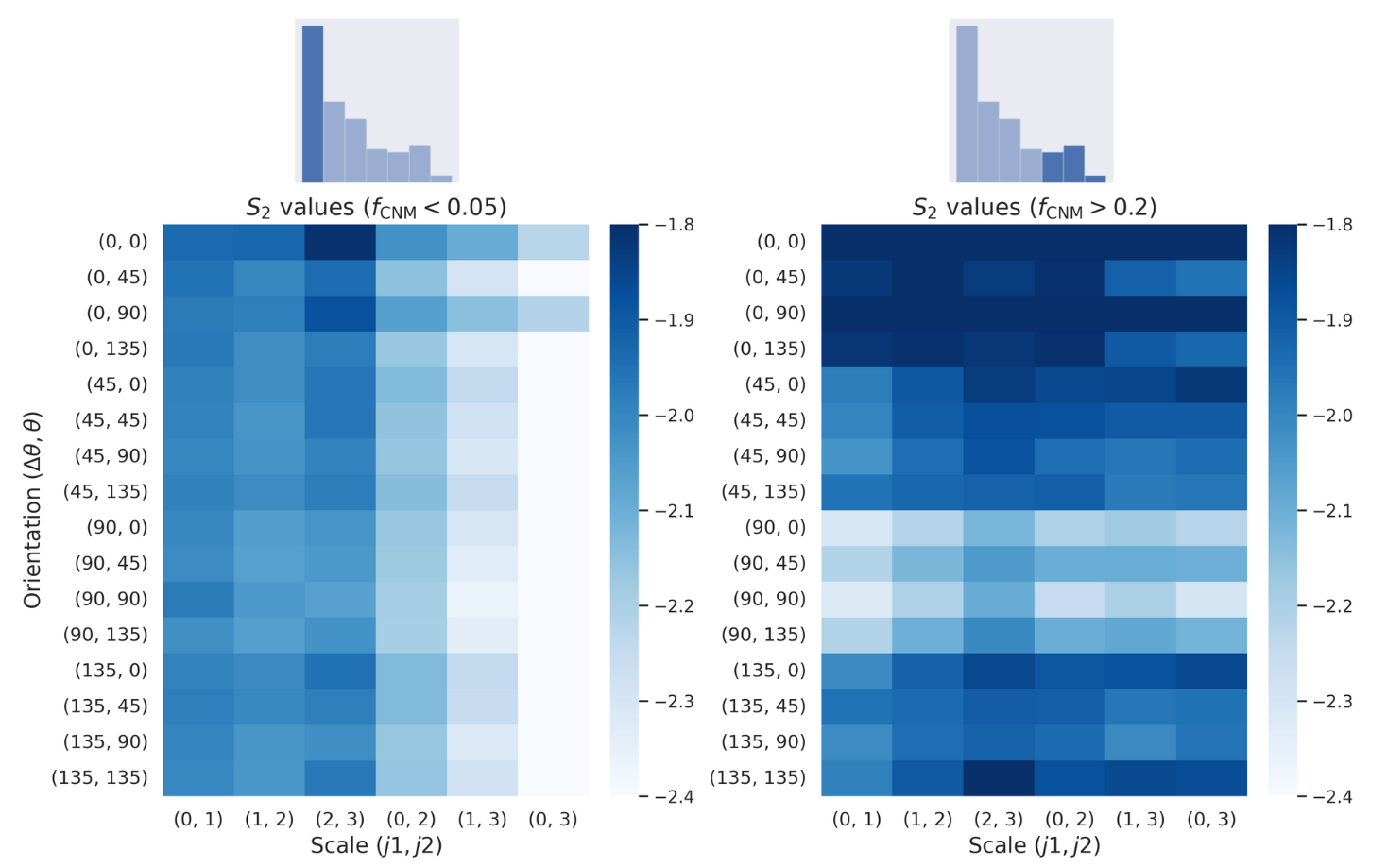}
        \caption{Left: the full $S_2(j_1, j_2, \Delta\theta, \theta)$ coefficients in log scale averaged over the 429 GALFA-\ion{H}{1} patches with $f_{\mathrm{CNM}}(v)<0.05$. Right: the $S_2(j_1, j_2, \Delta\theta, \theta)$ coefficients in log scale averaged over the 54 GALFA-\ion{H}{1} patches with $f_{\mathrm{CNM}}(v)>0.2$. The distributions of the $f_{\mathrm{CNM}}$ values are shown in log-scale histograms for each panel. The parallel components with $\Delta \theta = 0\degree$ are significantly larger in the large-$f_{\mathrm{CNM}}(v)$ case, while the small-scale $\Delta \theta = 90\degree$ components are much smaller for large-$f_{\mathrm{CNM}}(v)$ patches, corroborating the correlation of $f_{\mathrm{CNM}}$ with linear features.} \label{fig:st_val_fcnm}
    \end{minipage}
\end{figure*}

\begin{figure}[t]
\centering
\includegraphics[scale=0.52]{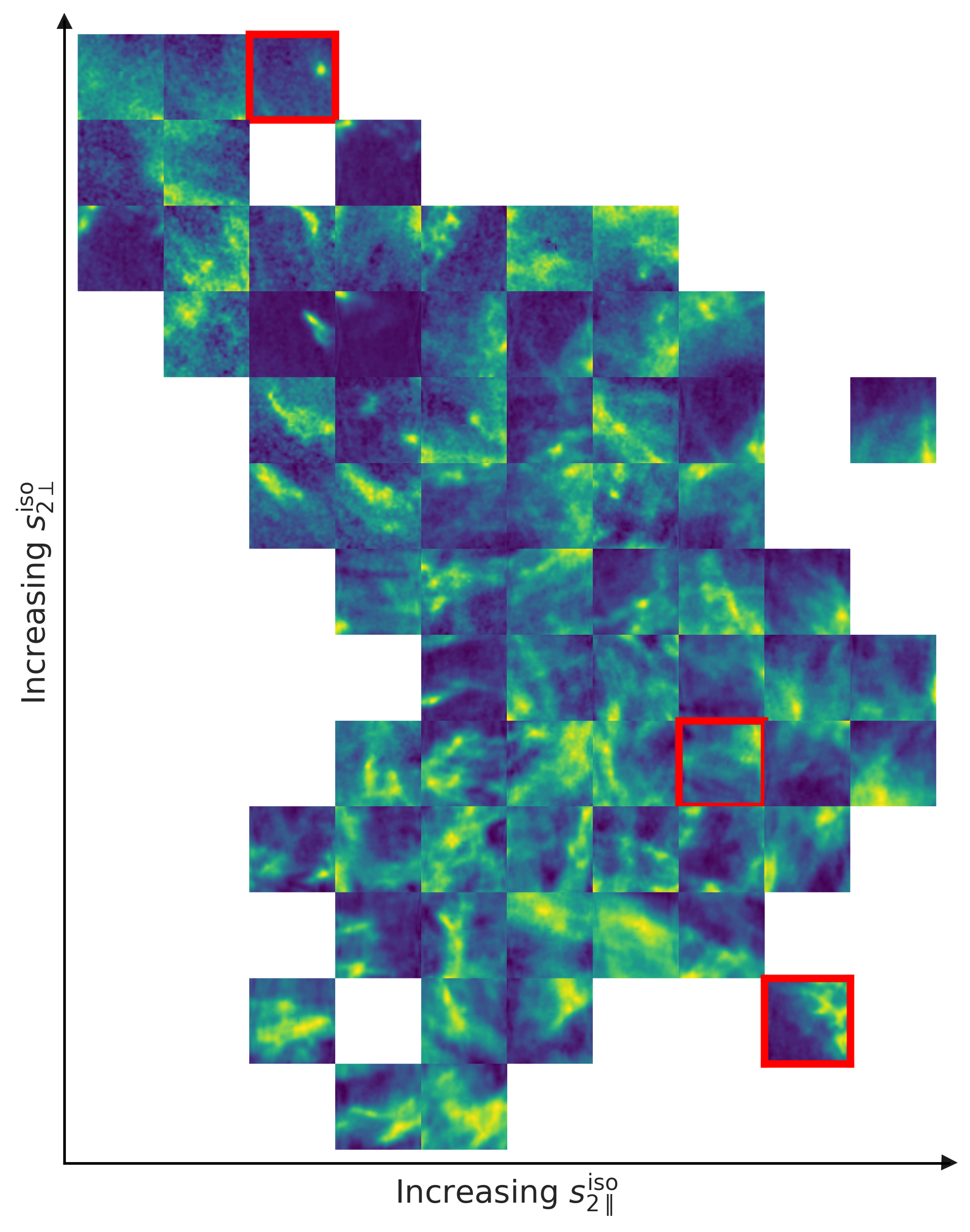}
\caption{Cluster of randomly selected GALFA-\ion{H}{1} patches, ordered by their small-scale $(j_1=0, j_2=1)$ $S^{\mathrm{iso}}_{2\parallel}$, $S^{\mathrm{iso}}_{2\perp}$ coefficient values. The GALFA-\ion{H}{1} images are $64'\times64'$ patches around the 58 sightlines with absorption measurements, as described in Section \ref{subsec:hi_abs}. The morphology of the patches highlighted in red are further examined in Figure \ref{fig:st_patch_compare}.}
\label{fig:st_cluster}
\end{figure}

\section{Morphology Exploration with the ST} \label{sec:hi_morph_res}
\subsection{Applying the ST to \ion{H}{1} Emission Data} \label{subsec:apply_st_data}
To motivate the use of ST morphology measures as a probe of the CNM content, in this section we examine the results of applying the ST to the \ion{H}{1} emission data and explore the interpretation of the resulting coefficients. We construct $N\times N$ pixel patches from GALFA-\ion{H}{1} maps, where each pixel spans $1\arcmin\times1\arcmin$, centered around the 30 sightlines with absorption measurements from 21-SPONGE described in Section \ref{subsec:hi_abs}. We construct a patch for each velocity channel in $|v_{\mathrm{LSR}}| < 30$ km/s with a channel width of $\delta v\sim3$ km/s, for a total of 570 channel maps. We compute the corresponding $f_{\mathrm{CNM}}(v)$ values from the emission/absorption pairs, as described in Section \ref{subsec:fcnm_data}, and adopt several preprocessing steps, before applying the ST to the GALFA-\ion{H}{1} patches. First, we interpolate over the background continuum source, using nearest-neighbor interpolation over the $3'\times3'$ region around the source at the center of each patch. Then we apply Fourier filtering to remove the fixed-angle patterns due to telescope scan artifacts \citep{Peek2017-ql}. The effect of this filtering is further discussed in Appendix \ref{appx:systematics}. Finally, to mitigate edge effects resulting from the boundaries of the square patches, we apply a circular apodization mask, tapered with a cosine function to each patch, with the apodization scale set to the patch size. We then apply the ST with scales $1\leq j < J$ and orientations $0 \leq l < L$ to the constructed GALFA-\ion{H}{1} patches to derive a set of ST coefficients per patch. 

For all of our ST calculations, we make use of the publicly available $\texttt{scattering}$ package introduced in \citet{cheng20-an}, which is based on the $\texttt{KYMATIO}$ package \citep{kymatio18_an}. In this study, we consider the full set of decorrelated and reordered second-order coefficients $S_2(j_1, j_2, \Delta\theta, \theta)$, which will further motivate the reductions when we examine their correlation with the CNM mass fraction $f_{\mathrm{CNM}}$. We choose the parameters: patch size $N=64\arcmin$ (pixels), scale $J=\log_2{N}-1=4$, and orientation $L=4$. Here, the choice of patch size is motivated by our goal of probing $f_{\mathrm{CNM}}$ for the sightline at the center of the patch. We want the patch size to be large enough to have enough spatial dynamic range, but not so large that it contains too much background information that is irrelevant to the CNM content of the central sightline. The upper limit of $J$ is set by the patch size $N$. In principle, $L$ can be arbitrarily large. However, increasing $L$ results in additional coefficients that are highly correlated with one another, since the number of independent coefficients is limited by the pixelization of the patches. We choose $L=4$ for our analysis, to balance encoding additional information with having a compact set of descriptors. 

\begin{figure}[t]
\centering
\includegraphics[scale=0.3]{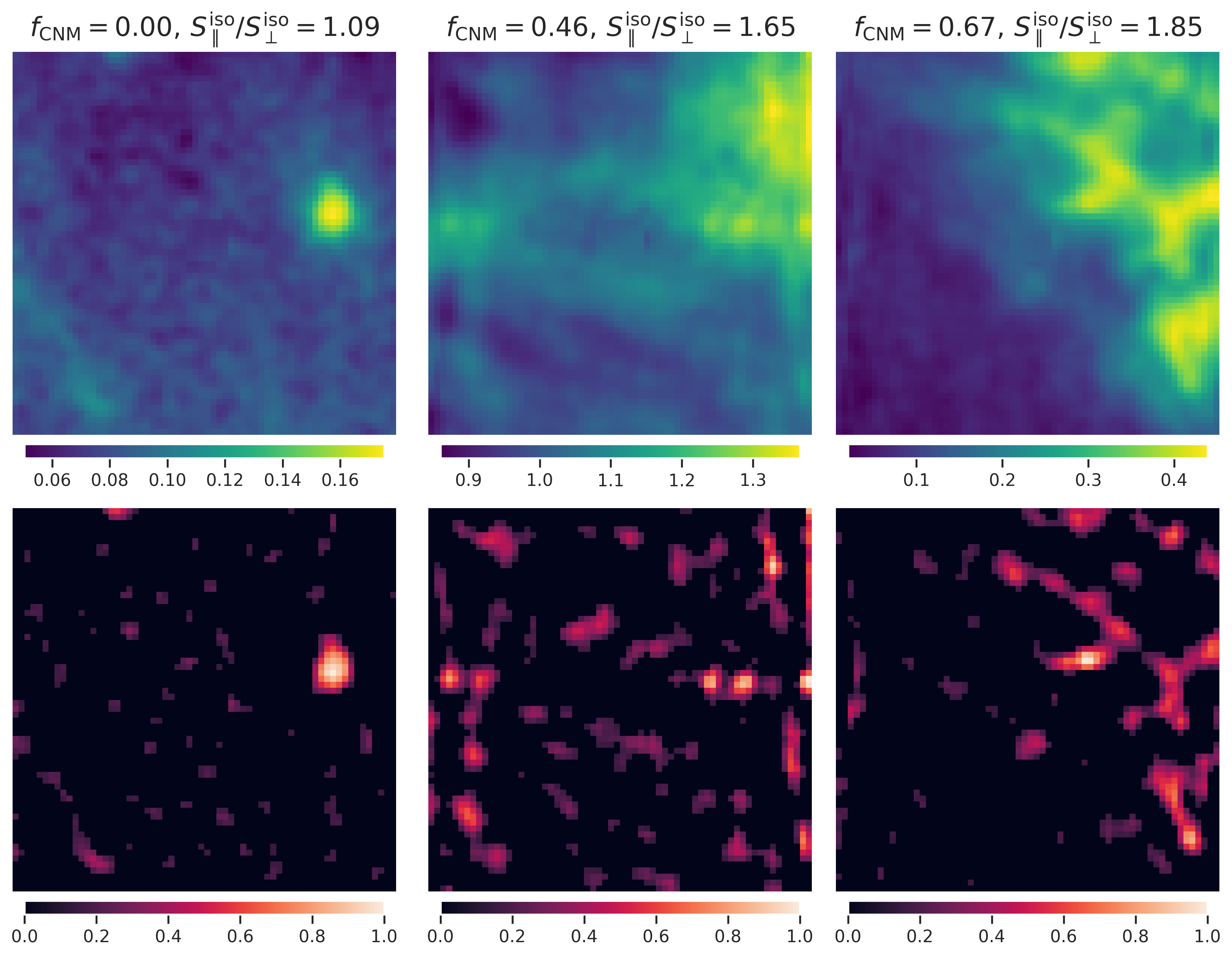}
\caption{Comparison of representative GALFA-\ion{H}{1} patches with differing values of $f_{\mathrm{CNM}}$ and small-scale $S_{\parallel}/S_{\perp}$ coefficients. Top: GALFA-\ion{H}{1} patches with color bars in units of $10^{20}\mathrm{cm}^{-2}$. Bottom: corresponding patches after applying the USM with radius $5'$. The patches with higher $S^{\mathrm{iso}}_{\parallel}/S^{\mathrm{iso}}_{\perp}$ and correspondingly higher $f_{\mathrm{CNM}}$ values contain features that are more coherent slender, and linear. The selected images are highlighted in red in Figure \ref{fig:st_cluster}. }
\label{fig:st_patch_compare}
\end{figure}

\begin{figure*}[t]
\centering
\includegraphics[width=0.82\textwidth]{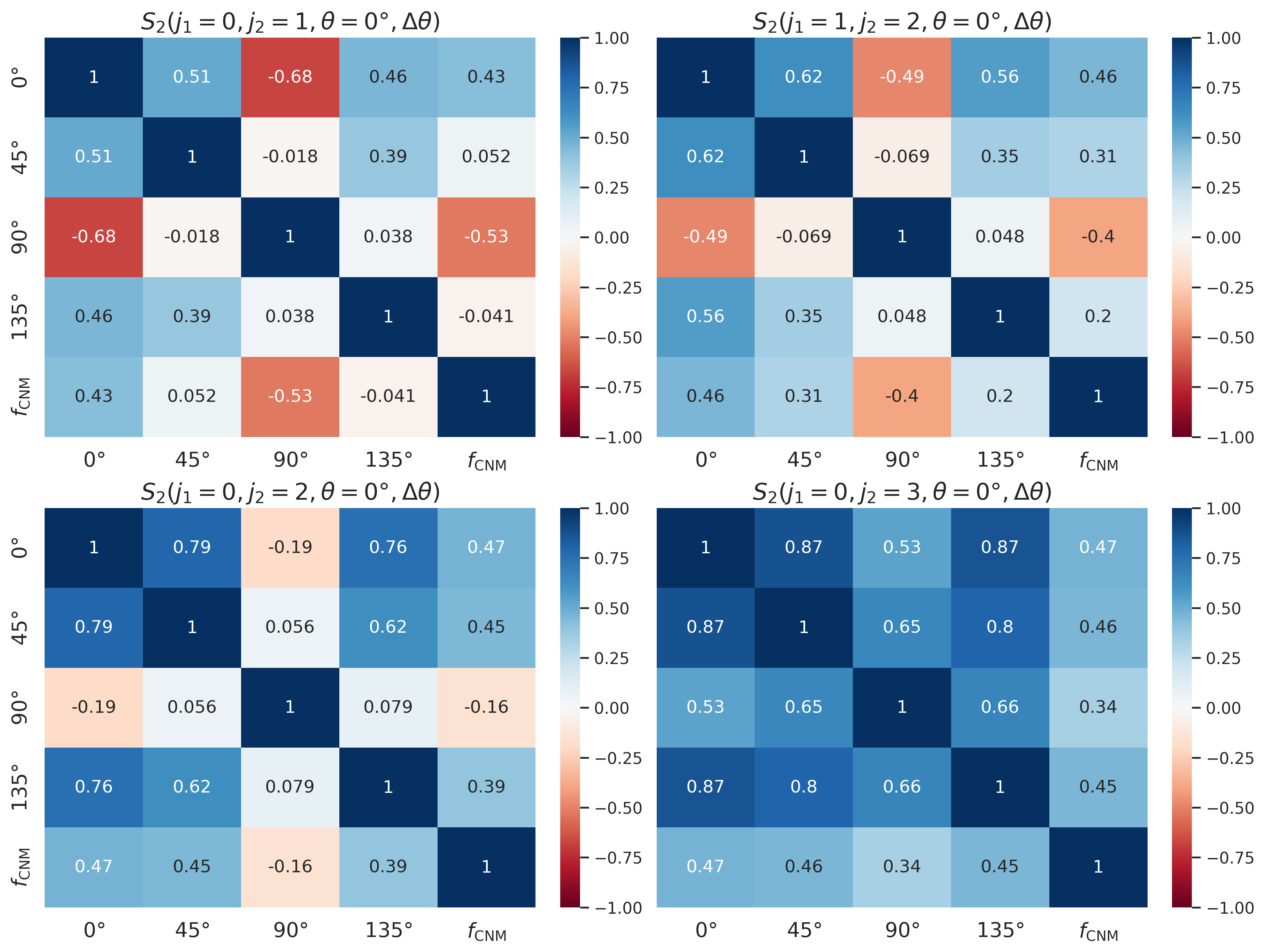}
\caption{Correlation matrix between the ST coefficients of different alignment angles $\Delta\theta=0\degree, 45\degree, 90\degree, 135\degree$, and $f_{\mathrm{CNM}}$, repeated in each subplot for different scale coefficients: $(j_1, j_2) = (0, 1), (1, 2), (0, 3), (0, 2)$, clockwise from top left. The parallel components with $\Delta\theta=0\degree$ consistently correlate with $f_{\mathrm{CNM}}$ across scales, while the perpendicular component $\Delta\theta=0\degree$ shows scale-dependent behavior, with strong anticorrelation at small scales, and a trend toward positive correlation at large scales.}
\label{fig:st_corr_orien}
\end{figure*}

\subsection{ST coefficients as \ion{H}{1} Morphology Measures} \label{subsec:apply_st_intp}
We apply the ST to emission patches centered at velocity $v$ to derive one set of coefficients per patch, then correlate them with the $f_{\mathrm{CNM}}(v)$ derived from the absorption measurements centered at that patch. The results are shown in Figure \ref{fig:st_val_corr}, where we present the full $S_2(j_1, j_2, \Delta\theta, \theta; v)$ coefficients averaged over all GALFA-\ion{H}{1} patches, along with the correlation of these coefficients with the $f_{\mathrm{CNM}}(v)$ values. The correlation metric used is the Spearman's rank correlation. The ST coefficient and correlation matrices are presented with the x-axis showing the scale components $(j_1, j_2)$, ordered by $|j_1-j_2|$, and the y-axis showing the orientation components $(\Delta\theta, \theta)$, ordered by $\Delta\theta = |l_1-l_2|$. For the plot of the average $S_2$ values, along the the scale dimension higher ST values are concentrated at smaller scales. Along the orientation dimension, the values are largely isotropic with respect to the absolute orientation angle $\theta$. This aligns with our expectation of not having a preferred absolute orientation for the \ion{H}{1} emission structures. With respect to the relative orientation $\Delta \theta$, the averaged parallel $\Delta \theta=0\degree$ coefficients are slightly larger than the perpendicular $\Delta\theta=90\degree$ coefficients, indicating that the general texture of the diffuse \ion{H}{1} in narrow-channel maps is more linear/filamentary than soft/blobby. 

The importance of the small-scale parallel and perpendicular coefficients is demonstrated more clearly in the correlation plot on the right, where we find a strong positive correlation between $f_{\mathrm{CNM}}$ and the parallel ST components at smaller scales, $(j_1, j_2) = (0,1), (1,2)$, and a strong anticorrelation with the perpendicular ST components at the same scales. Here, the scale parameter j translates to physical scale $2^{j+1}$ arcmin. If we adopt a fiducial distance to the \ion{H}{1} emission of 100 pc, $j=2$ corresponds to $8'$ or $\sim$0.2 pc. Given the discussion about the physical interpretations of these coefficients in Section \ref{subsec:st_reduc}, we can interpret the correlations as indicating that small-scale linear spatial structures in the vicinity of the absorption-measured sightlines are more prominent in channel maps with a higher LOS $f_{\mathrm{CNM}}$. A similar result is found in Figure \ref{fig:st_val_fcnm}, where we look at the values of the same ST coefficients, averaged over patches with small $f_{\mathrm{CNM}}$ values $\leq$ 0.05 vs. large $f_{\mathrm{CNM}}$ values $\geq$ 0.3. The patches with large $f_{\mathrm{CNM}}$ values have significantly larger parallel ($\theta=0\degree$) ST coefficients and smaller perpendicular ($\Delta\theta=90\degree$) coefficients at small scales. This behavior is what motivates our consideration of the reduced $S_{2\parallel}$ and $S_{2\perp}$ coefficients described in Section \ref{subsec:st_reduc}. 

In Figure \ref{fig:st_cluster} we show the clustering of GALFA-\ion{H}{1} patches by their small-scale $S_{2\parallel}^{\mathrm{iso}}$ and $S_{2\perp}^{\mathrm{iso}}$ coefficients. The small-scale coefficients organize the patches into coherent regions with distinct morphological features, and show a clear trend of anticorrelation, which is consistent with the correlation values between parallel and perpendicular ST coefficients in Figure \ref{fig:st_corr_orien}. The relationship between $f_{\mathrm{CNM}}$ and the ST morphology measures will be demonstrated more rigorously in the correlation studies presented in the following sections. In Figure \ref{fig:st_patch_compare} we select a few patches with representative values of small-scale ST coefficients and $f_{\mathrm{CNM}}$. To more clearly highlight the small-scale features, the unsharp mask (USM) filtered versions of these images are shown in the bottom panels, using a circular top-hat kernel with radius $5'$. The USM filters out low-frequency structures, by subtracting a smoothed version of an image from the original, then thresholding the filtered image at 0. It is visually apparent that the patches with higher $S_{\parallel}/S_{\perp}$ and correspondingly higher $f_{\mathrm{CNM}}$ values contain more coherent linear features.

Looking beyond the small-scale correlation patterns, the ST coefficient value and correlation matrices in Figure \ref{fig:st_val_corr} also show that the correlation of $f_{\mathrm{CNM}}$ with $S_{2\parallel}$ and  $S_{2\perp}$ is scale-dependent. This is especially true for the $S_{2\perp}$ coefficients, which are strongly anticorrelated with $f_{\mathrm{CNM}}$ at small scales $(j_1, j_2) = (0,1), (1,2)$, but show a weaker anticorrelation and even positive correlation toward larger scales. This suggests that the ratio $S_{\parallel}/S_{\perp}$ may not tell the full story: the full $S_{2\perp}$ coefficients could contain further $f_{\mathrm{CNM}}$-relevant morphological information that is not degenerate with $S_{2\parallel}$. Even though we would naively expect a linear feature to have large $S_{2\parallel}$ values and correspondingly small $S_{2\perp}$ values at a given scale, the full ST coefficients allow us to probe complex scale and orientation interactions that describe richer sets of morphological patterns. In Appendix \ref{appx:st_synthesis}, we present a more qualitative exploration, using synthesized images of the kind first described in Section \ref{subsec:st_reduc}. 

To further illustrate this interesting scale-dependent behavior, in Figure \ref{fig:st_corr_orien}, we show the correlation matrix between ST coefficients of different alignment angles $\Delta\theta=0\degree, 45\degree, 90\degree, 135\degree$, and $f_{\mathrm{CNM}}$, repeated for different scale components $(j_1, j_2) = (0, 1), (1, 2), (0, 2), (0, 3)$. The correlation coefficient between the parallel component $\Delta\theta=0\degree$ and the perpendicular component $\Delta\theta=90\degree$ ranges from -0.72 at scale $(0, 1)$ to +0.56 at scale $(0, 3)$, while the correlation of the perpendicular component with $f_{\mathrm{CNM}}$ changes correspondingly, from -0.7 to +0.38 at the same scales. Motivated by this scale-dependent correlation, we choose $S_{2\parallel}^{\mathrm{iso}}$ and $S_{2\perp}^{\mathrm{iso}}$ as the lowest-order coefficients to be considered, instead of further reducing these quantities to their ratio $S_{2\parallel}^{\mathrm{iso}}/S_{2\perp}^{\mathrm{iso}}$. That ratio is a good measure of linear features at small scales, but correlates less well with $f_{\mathrm{CNM}}$ at large scales, where $S_{2\parallel}^{\mathrm{iso}}$ and $S_{2\perp}^{\mathrm{iso}}$ are positively correlated. 

In summary, the morphology of the GALFA-\ion{H}{1} patches, as probed by the ST morphology measures, shows a clear trend of small-scale linear features being highly correlated with CNM content. Here and in subsequent discussions, ``small-scale" specifically denotes the ST scale parameters $(j_1, j_2) = (0, 1), (1, 2)$, where the scale parameter $j$ translates to a physical scale of $2^{j+1}$ arcmin. These correlations are consistent with past results showing that small-scale \ion{H}{1} intensity structures are preferentially CNM \citep{clark19_pn}. This will be further examined and quantified in the following section. Additionally, beyond small scales, the correlations also show interesting scale-dependent behavior, where $S_{\perp}$ is anticorrelated with $f_{\mathrm{CNM}}$ at small scales, but positively correlated at large scales, indicating that the CNM is potentially associated with additional morphologies beyond the most prominent small-scale filamentary features. 

\begin{figure*}[!t]
\centering
\includegraphics[width=\textwidth]{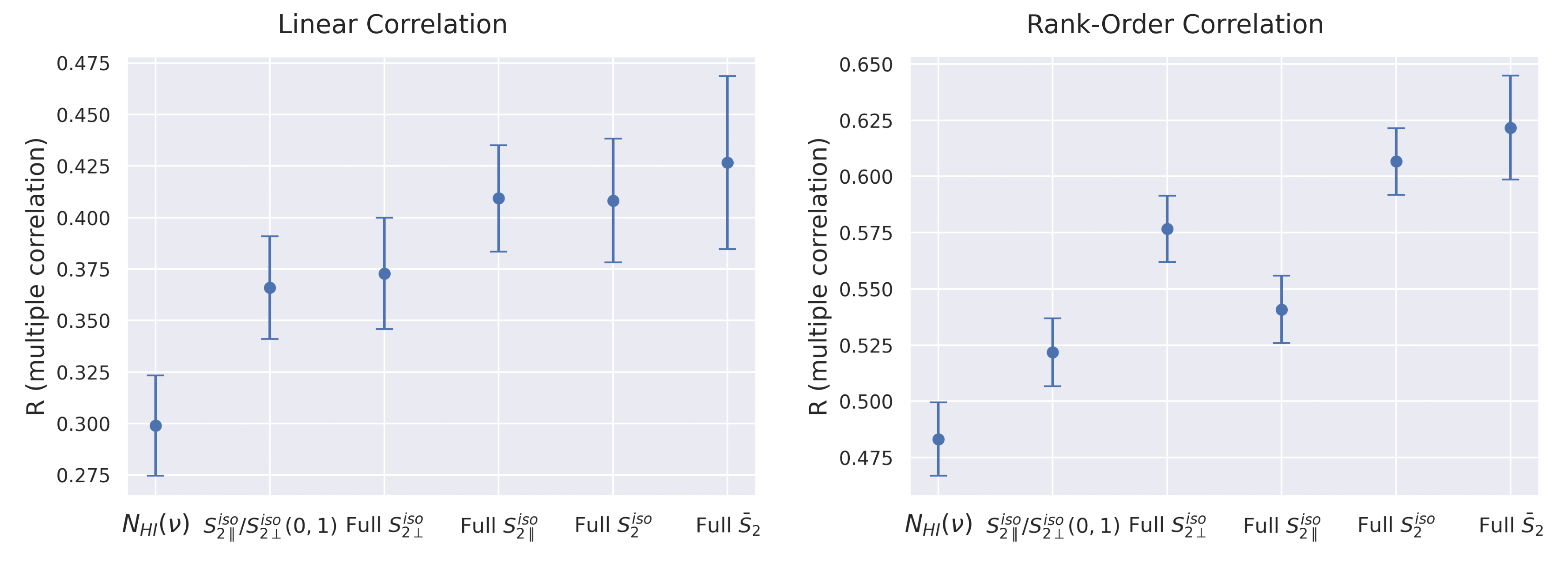}
\caption{Comparison of the per-channel $f_{\text{CNM}}(v)$ correlation between various sets of ST coefficients and the mean column density $N_{\text{H\,I}}(v)$. The corrected coefficient of multiple correlation \citep{abdi07_mc} is used as the metric. The left panel shows the standard linear version, while the right panel shows the rank order version, where the correlation is computed over the Spearman rank of the variables. The ST coefficients are ordered by the degree of reduction, so that the total number of coefficients increases from left to right. Accounting for uncertainty, all versions of the ST coefficients are more predictive of $f_{\mathrm{CNM}}$ than $N_{\mathrm{H\,I}}$, including the case using a single parameter at scale $(j_1, j_2)=(0, 1)$.}
\label{fig:multi_corr_v}
\end{figure*}

\section{CNM Correlation Studies with the ST} \label{sec:fcnm_corr_res}
In the previous section, we explored the use of ST coefficients as morphology measures of the \ion{H}{1} emission, and the interpretation of CNM-correlated ST coefficients. In this section, we study correlations between the sequence of ST coefficients introduced in Equations \ref{eq:st_full}--\ref{eq:st_po_iso} and $f_{\mathrm{CNM}}$, and demonstrate the potential for inferring the CNM content of interstellar gas from the spatial structure of the \ion{H}{1} emission. This study examines two sets of $f_{\mathrm{CNM}}$ data, both computed from absorption/emission spectral pairs along the 58 21-SPONGE/Millennium sightlines described in Section \ref{subsec:fcnm_data}. First, we look at $f_{\mathrm{CNM}}(v)$ over the narrow velocity channels with $\delta v = 3$ km/s for the 30 21-SPONGE sightlines with high optical depth sensitivities, and high-velocity resolution. We apply the ST to GALFA-\ion{H}{1} patches in channel maps of the same (3 km/s) velocity width, to derive a spectrum of ST coefficients $S_2(j_1, j_2, \Delta\theta, \theta; v)$ per sightline to compare with the corresponding $f_{\mathrm{CNM}}(v)$ spectra. We then examine the correlations with the full velocity-integrated LOS $f_{\mathrm{CNM}}$ along the full 58 21-SPONGE/Millennium sightlines. The process of constructing GALFA-\ion{H}{1} patches around the sightline and applying the ST in this case will be described in more detail in Section \ref{subsec:corr_fcnm_i}. Finally, in Section \ref{subsec:fir_nhi}, we look at FIR/$N_{\mathrm{H\,I}}$ histograms, binned by ST morphology measures, over the full GALFA-\ion{H}{1} sky, as an additional test of the correlation of these coefficients with cold gas content, independent of the $f_{\mathrm{CNM}}$ data from absorption+emission measurements. 

\begin{figure*}[t]
\centering
\includegraphics[width=0.9\textwidth]{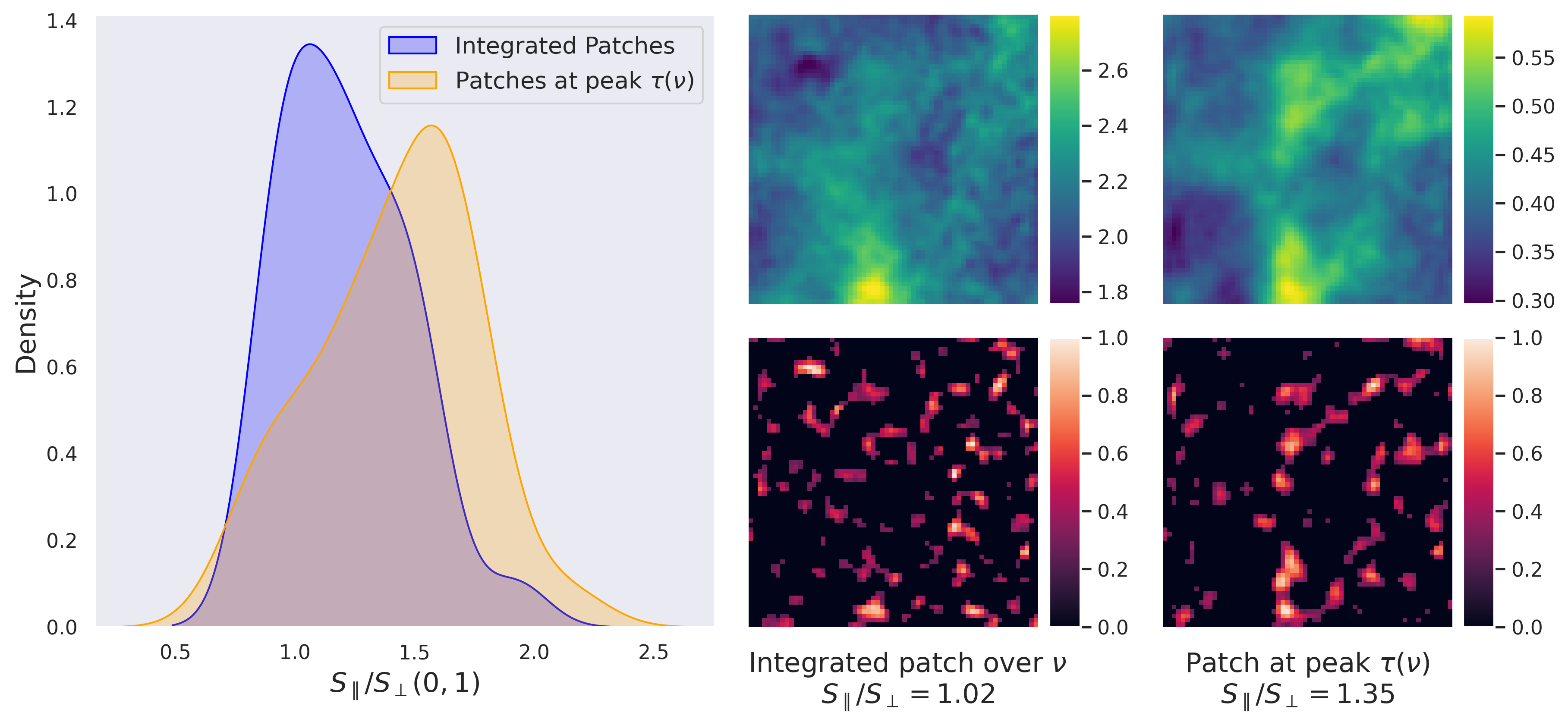}
\caption{Comparing the morphology of integrated vs. narrow-channel GALFA-\ion{H}{1} patches, as characterized by the small-scale linear ST coefficient $S_{\parallel}/S_{\perp}(0, 1)$. The left panel shows the kernel density estimate plot of the ST coefficients for patches around the 58 21-SPONGE/Millenium sightlines. There is a shift toward higher $S_{\parallel}/S_{\perp}(0, 1)$ values when the ST is applied to patches that are taken at peak $\tau(v)$ over narrow width $\Delta v\sim3$km/s, compared to patches integrated over the full velocity range $|v|<40$ km/s. This is consistent with preferentially CNM small-scale linear structures being more prominent in narrow channels \citep{clark14-ma, clark19_pn}. The result is further illustrated in the right panels, for a sample sightline, 3C245A. These panels show the total brightness temperature (K; top) and the normalized USM emission (bottom) for velocity-integrated (left) and peak $\tau(v)$ (right) patches. The narrow-channel patch shows more prominent and coherent linear features being highlighted by the USM, and has a correspondingly larger $S_{\parallel}/S_{\perp}(0, 1)$ coefficient.}
\label{fig:vpeak_patch}
\end{figure*}

\begin{figure}[t]
\centering
\includegraphics[scale=0.52]{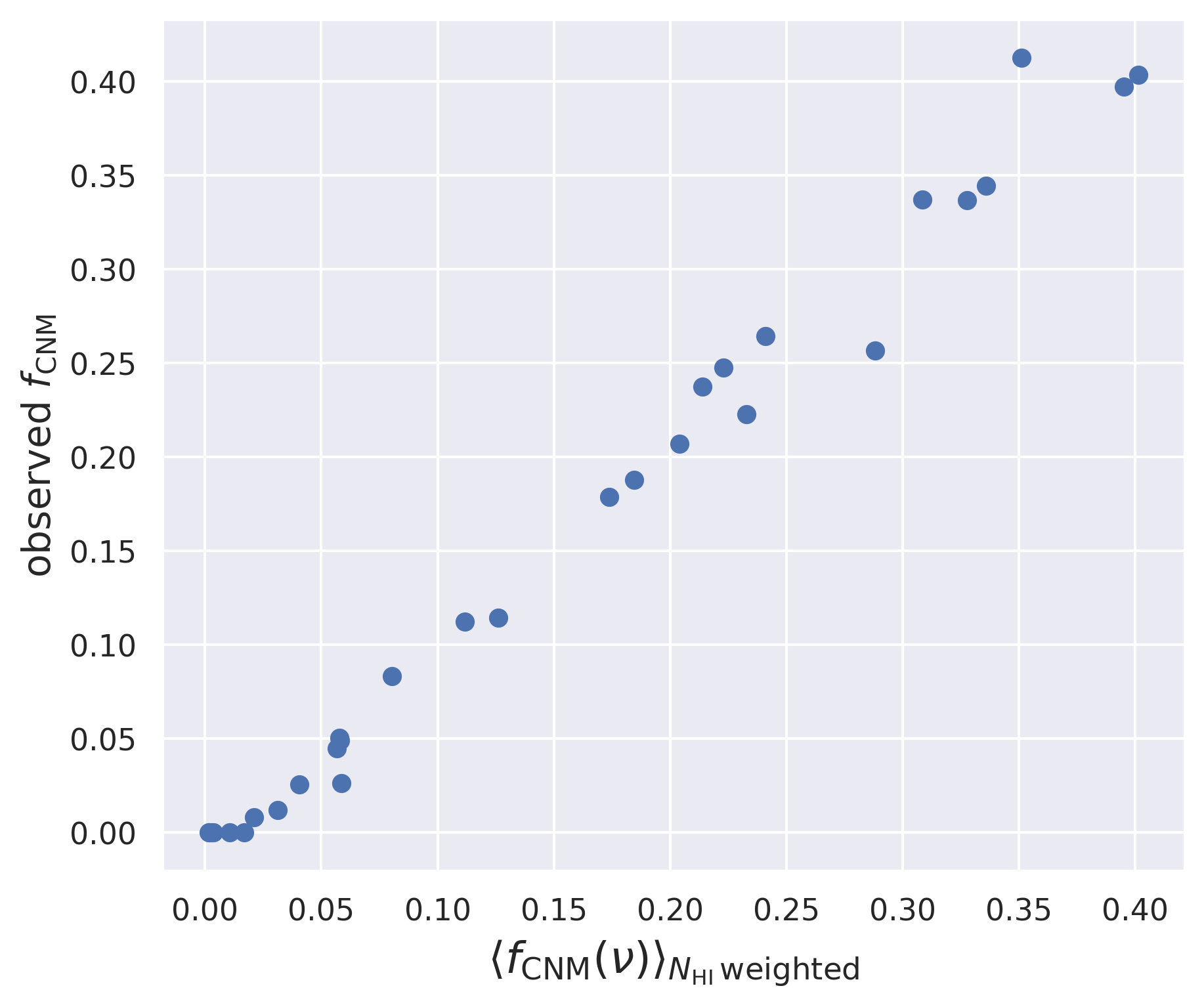}
\caption{Evaluating the performance of estimating the LOS $f_{\mathrm{CNM}}$ data from the $N_{\mathrm{H\,I}}$-weighted average of per-channel $f_{\mathrm{CNM}}(v)$ spectra. The good agreement shown here between the estimated $f_{\mathrm{CNM}}$ and the data lends confidence to the approach of deriving integrated ST measures from the ST being applied to per-channel data, as discussed in Section \ref{subsec:corr_fcnm_i}.}
\label{fig:fcnm_estimate}
\end{figure}

\begin{figure*}[t]
\centering
\includegraphics[width=\textwidth]{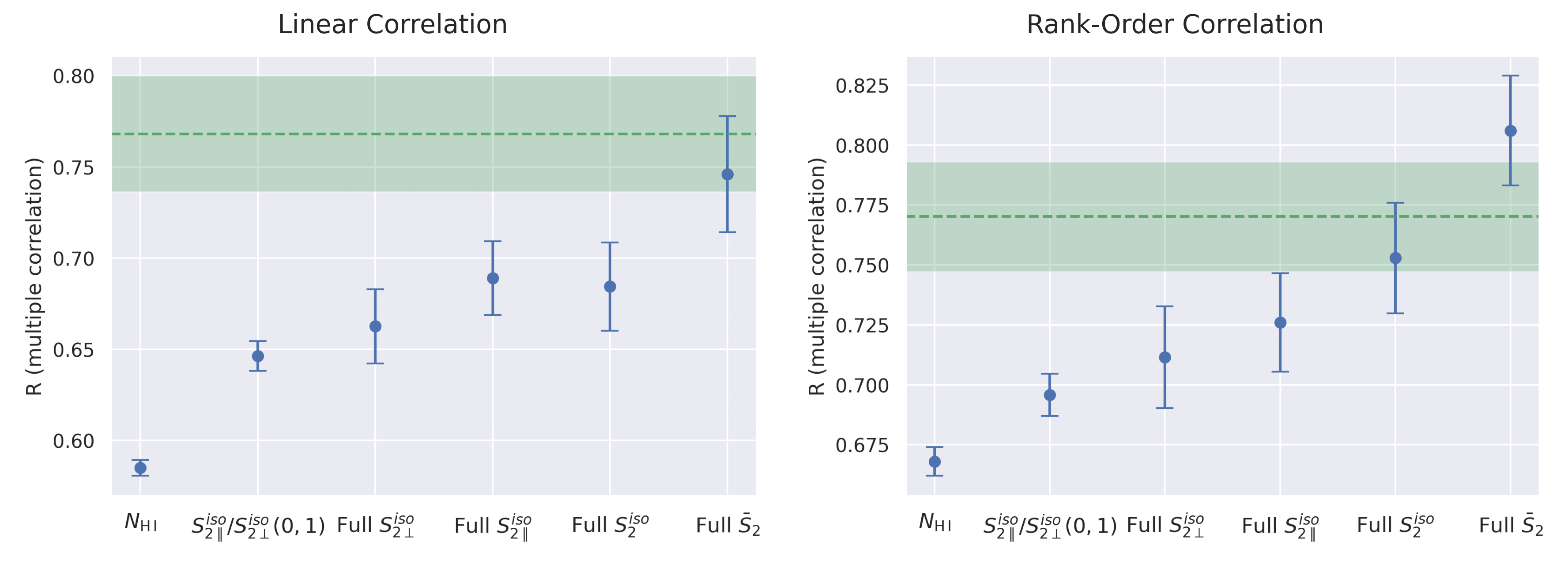}
\caption{Comparing the total LOS $f_{\text{CNM}}$ correlation performance of various sets of ST coefficients and the mean $N_{\text{HI}}$. The multiple-correlation coefficient \citep{abdi07_mc} is used as the metric, and a total of 58 sightlines with absorption measurements from 21-SPONGE and the Millennium survey are included in the analysis. Similar to the narrow velocity channel version described in Figure \ref{fig:multi_corr_v}, the different sets of ST coefficients all contain more $f_{\text{CNM}}$-correlating information beyond the $N_{\text{H\,I}}$ correlation. The overall correlation is higher than the per-channel version, likely due to the lower uncertainties than in the narrow-channel $f_{\text{CNM}}(v)$ estimation. The region highlighted in green is the test set performance of the CNN model introduced in \citet{Murray2020-uf}, where a simple Pearson/Spearman correlation is computed between the CNN prediction and the $f_{\mathrm{CNM}}$ measurements.}
\label{fig:multi_corr_los}
\end{figure*}

\subsection{Narrow velocity Channel $f_{\text{CNM}}(v)$ correlation} \label{subsec:corr_fcnm_v}
Motivated by results showing that linear features that are preferentially associated with the CNM are most prominent in narrow ($\sim$a few km/s) velocity channels \citep{clark19_pn}, in this section we examine the correlation between the ST morphology measures applied to narrow-channel \ion{H}{1} patches and $f_{\mathrm{CNM}}(v)$. The choice of channel width here is mostly limited by the optical depth sensitivity and velocity resolution. Thus, to get the most robust $f_{\mathrm{CNM}}(v)$ estimation, we restrict in this study to the high-sensitivity 21-SPONGE dataset, and velocity range $|v|<30$ km/s with channel width $\delta v = 3$ km/s. Narrower channel width ($\delta v = 1.5$ km/s) maps are also examined, producing qualitatively similar results, but with higher uncertainty and more outliers in the estimated $f_{\mathrm{CNM}}(v)$ distribution. The full 21-SPONGE/Millennium dataset over the range $|v|<40$ km/s will be considered in the next section, when we examine the velocity-integrated LOS $f_{\mathrm{CNM}}$. Examples of the estimated $f_{\mathrm{CNM}}(v)$ with its corresponding optical depth spectra $\tau(v)$ for two sample sightlines are shown in Figure \ref{fig:fcnm_spec}. A total of 570 $f_{\mathrm{CNM}}(v)$ measurements result from the 30 21-SPONGE sightlines with $\delta v = 3$ km/s over the range $|v|<30$ km/s. We compare the $f_{\mathrm{CNM}}(v)$ spectra per sightline to the ST coefficients $S_2(j_1, j_2, \Delta\theta, \theta; v)$ spectra that describe the spatial morphology in the vicinity of the sightline. To evaluate and compare the correlation performance of different sets of ST coefficients described in Equations \ref{eq:st_full}- \ref{eq:st_po_iso} with the $f_{\mathrm{CNM}}(v)$ data, we use the multiple-correlation coefficient \citep{abdi07_mc}, which generalizes the standard Pearson correlation coefficient to the case of multiple predictive variables. The coefficient of multiple correlation is given by
\begin{equation} \label{eq:multi_corr}
    R^2 = \boldsymbol{y}^\intercal \boldsymbol{x}^{-1} \boldsymbol{y},
\end{equation}
where $\boldsymbol{x}$ is the correlation matrix between the predictive variables and $\boldsymbol{y}$ is the correlation vector between the predictive variables and the target variable. To get a better estimate of the population statistic, which takes into account the sample size and number of predictive variables used, we adopt the corrected version of the multiple-correlation coefficient \citep{abdi07_mc}:
\begin{equation} \label{eq:multi_corr_correct}
    \tilde{R}^2 = 1 - \left[(1-R^2)\frac{N-1}{N-K-1}\right],
\end{equation}
where $R$ is the uncorrected coefficient, $N$ is the population size (the number of $f_{\mathrm{CNM}}$ measurements), and $K$ is the number of predictive variables (the number of ST coefficients) used to estimate the target variable. We also derive a Spearman rank order version of the multiple-correlation coefficient, by rank transforming the data and then computing Equations \ref{eq:multi_corr}-\ref{eq:multi_corr_correct} on the ranks. The resulting correlation values with uncertainties are shown in Figure \ref{fig:multi_corr_v}. The uncertainties are estimated from the target and predictive variable errors, through a Monte Carlo error propagation procedure. To estimate the uncertainty on the ST coefficients, we recompute the coefficients after adding a ``pure noise" component to the GALFA-\ion{H}{1} patches. The noise patches are constructed from GALFA-\ion{H}{1} emission in a velocity channel centered at $v=400$ km/s, with the same channel width $\delta v=3$ km/s. A more detailed discussion and results of the uncertainty quantification for the ST coefficients can be found in Appendix \ref{appx:systematics}. In Figure \ref{fig:multi_corr_v}, the following sets of ST parameters are compared in terms of their correlation performance with $f_{\mathrm{CNM}}$:
\begin{itemize}
    \item $S_{2\parallel}^{\mathrm{iso}}/S_{2\perp}^{\mathrm{iso}}(j_1=0, j_2=1)$;
    \item full $S_{2\perp}^{\mathrm{iso}}(j_1, j_2)$ coefficients;
    \item full $S_{2\parallel}^{\mathrm{iso}}(j_1, j_2)$ coefficients;
    \item full $S_2^{\mathrm{iso}}(j_1, j_2, \Delta\theta)$ coefficients; and
    \item full $S_2(j_1, j_2, \Delta\theta, \theta)$ coefficients.
\end{itemize}
where the coefficients are ordered by the degree of reduction applied, from a single coefficient per image $S_{2\parallel}^{\mathrm{iso}}/S_{2\perp}^{\mathrm{iso}}(j_1=0, j_2=1)$ to the full decorrelated $S_2(j_1, j_2, \Delta\theta, \theta)$ coefficients. In the case of a single coefficient, the multiple-correlation coefficient reduces to the standard Pearson correlation. As discussed in Section \ref{subsec:apply_st_intp}, at small scale $(j_1, j_2)=(0, 1)$, $S_{2\parallel}^{\mathrm{iso}}$ is strongly correlated with $f_{\mathrm{CNM}}$, while $S_{2\perp}^{\mathrm{iso}}$ is strongly anticorrelated, making $S_{2\parallel}^{\mathrm{iso}}/S_{2\perp}^{\mathrm{iso}}$ a good measure of the linear features that are evidently predictive of $f_{\mathrm{CNM}}$ at small scales. The $f_{\mathrm{CNM}}(v)$ correlations of the ST coefficients are further compared to the correlations of the mean column density in each velocity channel $N_{\mathrm{H\,I}}(v)$. $N_{\mathrm{H\,I}}$ is computed from the brightness temperature $T_b(v)$ under the optically thin assumption, a good approximation at high Galactic latitudes \citep{Murray2018-xe}. As the figure shows, the ST coefficients are more predictive of $f_{\mathrm{CNM}}$ than $N_{\mathrm{H\,I}}$ accounting for uncertainty, even in the case of the single-coefficient ST estimator. Comparing between the ST coefficients, the full $S_{2\perp}^{\mathrm{iso}}(j_1, j_2)$ contains more $f_{\mathrm{CNM}}$-correlating information than just the small-scale $(j_1, j_2)=(0, 1)$ $S_{2\parallel}^{\mathrm{iso}}/S_{2\perp}^{\mathrm{iso}}$ coefficient, while the full $S_2(j_1, j_2, \Delta\theta, \theta)$ further outperforms the full parallel coefficients. Identifying the additional morphological descriptors beyond small-scale linearity that correlate with $f_{\mathrm{CNM}}$ is an interesting question for future work. One possible contribution is related to the scale-dependent behavior discussed in Section \ref{subsec:apply_st_intp} and qualitatively explored in Appendix \ref{appx:st_synthesis}. 

\subsection{Velocity-integrated LOS $f_{\text{CNM}}$ correlation} \label{subsec:corr_fcnm_i}
We also carry out the same correlation study for the total LOS $f_{\mathrm{CNM}}$, integrated over the full velocity range $|v_{\mathrm{LSR}}| < 40$ km/s of the constructed spectral pairs, resulting in 58 data points from the 21-SPONGE/Millenium absorption measurement sightlines. We explore different approaches to the process of applying the ST to the GALFA-\ion{H}{1} patches in this LOS-integrated estimation. First, the ST can be directly applied to patches integrated over the full velocity range $|v_{\mathrm{LSR}}| < 40$ km/s. However, integrating over the full range stacks and blends the potentially most relevant morphological features. We illustrate this in Figure \ref{fig:vpeak_patch} by comparing the results of applying the ST to integrated versus narrow-channel patches. The narrow-channel patches are constructed with a channel width $\Delta v\sim3$ km/s around the peak $\tau(v)$ for each of the sightlines. The distribution of the small-scale $S_{\parallel}/S_{\perp}$ coefficients shifts toward higher values for the narrow-channel patches, compared to the velocity-integrated patches. This is consistent with the picture that preferentially CNM small-scale linear structures are more prominent in narrow channels \citep{clark14-ma, clark19_pn}, and suggests that we should derive a set of integrated measures from the ST coefficients applied to narrow-channel patches, instead of applying the ST directly to LOS-integrated patches. Motivated by our goal of constructing a point statistic for estimating a single value per sightline, while still utilizing the richer per-channel information, we define the following LOS-averaged ST coefficients weighted by the per-channel column density:
\begin{equation} \label{eq:nhi_weight_st}
    \left<S\right> = \frac{\int_{\mathrm{LOS}} S(v)*\mathrm{N_{H\,I}}(v)\,dv}{\int_{\mathrm{LOS}} \mathrm{N_{H\,I}}(v)\,dv}
\end{equation}
where $\left<S\right>$ is the integrated ST coefficient and $S(v)$ contains the narrow velocity channel ST coefficients, computed for each channel map along the same sightline. The weighting by mean column density is motivated by our goal of estimating $f_{\mathrm{CNM}}$, a mass fraction that is proportional to the column density. Before applying it to the ST coefficients, we validate this simple approximation using $f_{\mathrm{CNM}}$ data, by estimating the total LOS-integrated $f_{\mathrm{CNM}}$ from the per-channel $f_{\mathrm{CNM}}(v)$:
\begin{equation} \label{eq:nhi_weight_fcnm}
    \left<f_{\mathrm{CNM}}\right> = \frac{\int_{\mathrm{LOS}} f_{\mathrm{CNM}}(v)*\mathrm{N_{H\,I}}(v)\,dv}{\int_{\mathrm{LOS}} \mathrm{N_{H\,I}}(v)\,dv}.
\end{equation}
The results are shown \ref{fig:fcnm_estimate} with excellent agreement between this approximation and the $f_{\mathrm{CNM}}$ data derived from the full absorption/emission spectra. Thus, while the above equations are only simple approximations, since $f_{\mathrm{CNM}}(v)$ is not an additive quantity, and ${\int_{\mathrm{LOS}} \mathrm{N_{H\,I}}(v)\,dv}$ makes the optically thin assumption less accurate for high $f_{\mathrm{CNM}}$ regions, this is a reliable way of deriving an integrated measure from per-channel data and an improvement upon applying the ST directly to velocity-integrated GALFA-\ion{H}{1} patches. 

Using this approach, we conduct a multiple-correlation study as described in the previous section, and present the results in Figure \ref{fig:multi_corr_los}. The plot is presented similarly to the one in Figure \ref{fig:multi_corr_v}, where sets of ST coefficients, ordered by level of reduction, are compared in terms of their $f_{\mathrm{CNM}}$ correlation with one another and with the mean column density $N_{\mathrm{H\,I}}$ integrated over the same velocity range as the $f_{\mathrm{CNM}}$ data. The correlation values are higher than those of the per-channel version in Figure \ref{fig:multi_corr_v}, likely due to the higher uncertainty in the narrow-channel $f_{\text{CNM}}(v)$ estimation. We observe similar qualitative results as in the previous section. The ST coefficients are more predictive of $f_{\mathrm{CNM}}$ than the column density, even in the case of a single coefficient $S_{2\parallel}^{\mathrm{iso}}/S_{2\perp}^{\mathrm{iso}}(0, 1)$, which is a measure of small-scale linear structures. Significant additional $f_{\mathrm{CNM}}$-correlating information is found in the full ST coefficients. 

In Figure \ref{fig:multi_corr_los}, we also show in green bands the performance of the CNN model's $f_{\mathrm{CNM}}$ prediction \citep{Murray2020-uf}, where a simple Pearson/Spearman correlation is computed between the CNN prediction and the $f_{\mathrm{CNM}}$ data. Note that in the case of the CNN model, the CNN is trained on simulation data, then independently validated on the $f_{\mathrm{CNM}}$ dataset. By contrast, our results constitute a model-independent correlation study of the ST coefficients as predictive variables of $f_{\mathrm{CNM}}$. The takeaway from Figure \ref{fig:multi_corr_los} is that the ST coefficients derived from the \ion{H}{1} emission morphology potentially contain comparable $f_{\mathrm{CNM}}$-correlating information to the spectral information extracted by the CNN model.

\begin{figure*}[t]
\centering
\includegraphics[width=0.8\textwidth]{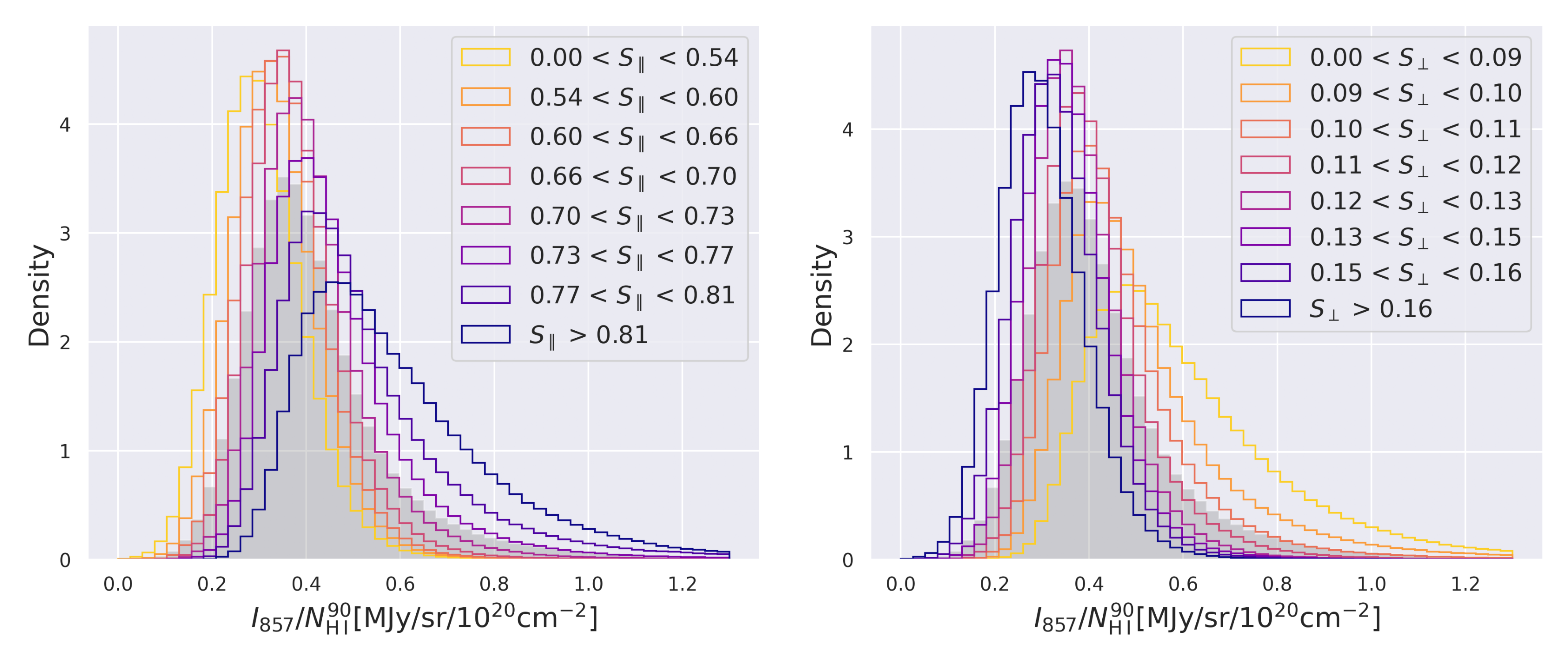}
\caption{Histograms of $I_{857}/N_{\mathrm{H\,I}}$ for the high-Galactic latitude ($|b|>30\degree$) GALFA-\ion{H}{1} sky, partitioned by the ST coefficients $S_\parallel$ and $S_\perp$ into bins with equal numbers of sightlines. $S_\parallel\equiv S_{2\parallel}^{\mathrm{iso}}(j_1=0, j_2=1)$ and $S_\perp\equiv S_{2\parallel}^{\mathrm{iso}}(j_1=0, j_2=1)$ respectively, converted to log scale and normalized to [0, 1]. The $I_{857}/N_{\mathrm{H\,I}}$ histogram for the full sky is plotted in gray. }
\label{fig:fn_hist}
\end{figure*}

\begin{figure*}[!ht]
\centering
\includegraphics[width=\textwidth]{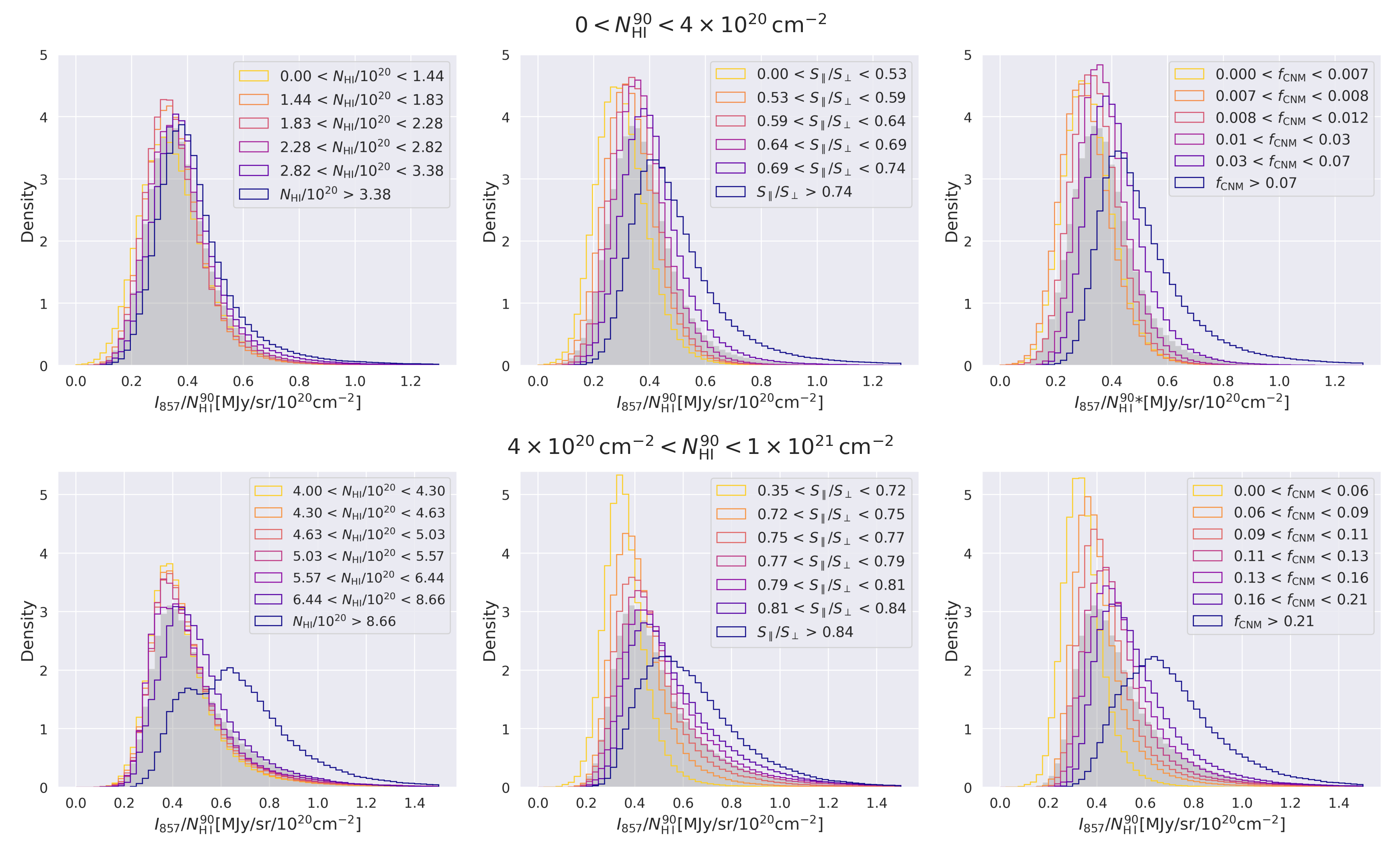}
\caption{Histograms of $I_{857}/N_{\text{HI}}$ for the high-Galactic latitude ($|b|>30\degree$) GALFA-\ion{H}{1} sky, partitioned by $N_{\mathrm{H\,I}}$, $S_\parallel/S_\perp$, and the CNN-predicted $f_{\mathrm{CNM}}$ \citep{Murray2020-uf}. The histogram bins are defined such that each bin has an equal number of sightlines. We repeat this analysis for different column density regimes, with the $I_{857}/N_{\text{H\,I}}$ histogram for the full sky in the regime plotted in gray. Top: the lower column density regime, with $N_{\mathrm{H\,I}}<4\times 10^{20}\,\mathrm{cm}^{-2}$; Bottom: the higher column density regime - $4\times 10^{20}\,\mathrm{cm}^{-2} < N_{\mathrm{H\,I}} < 1\times 10^{21}\,\mathrm{cm}^{-2}$. }
\label{fig:fn_hist_regimes}
\end{figure*}

\subsection{FIR/$N_{\mathrm{H\,I}}$ ratio} \label{subsec:fir_nhi}

Our model-independent correlation studies using $f_{\mathrm{CNM}}$ data from absorption measurements show that the \ion{H}{1} morphology information extracted by the ST is highly predictive of the CNM content, and potentially contains comparable $f_{\mathrm{CNM}}$-correlating information to the spectral information used by traditional methods of phase decomposition. Specifically, the ST components that are indicative of linear features at small scales are by themselves highly correlated with $f_{\mathrm{CNM}}$. In this section, we further examine the correlation of these small-scale components with the cold gas content independent of the absorption measurement data, by looking at the FIR/\ion{H}{1} column ratio binned by these coefficients. The FIR emission data come from the all-sky map at 857 GHz from Planck \citep{planck18-I}. The \ion{H}{1} column density data are from the stray radiation-corrected GALFA-\ion{H}{1} column density map constructed from the \ion{H}{1} intensity integrated over $|v_{\mathrm{LSR}}|\leq90$ km/s \citep{Peek2017-ql}. To distinguish this map from the column density in Section \ref{subsec:corr_fcnm_i} integrated over the same velocity range as the $f_{\mathrm{CNM}}$ data, we denote the $|v_{\mathrm{LSR}}|\leq90$ km/s version here as $N_{\mathrm{H\,I}}^{90}$. 

The distribution of the FIR/$N_{\mathrm{H\,I}}$ ratio was examined in \citet{clark19_pn}, who found that small-scale magnetically aligned linear features in \ion{H}{1} channel maps are real density structures. That work observed an enhancement of FIR/$N_{\mathrm{H\,I}}^{90}$ for regions of the sky with higher measures of small-scale linear intensity. The physical effects that are expected to raise this ratio are all associated with increased cold-phase content \citep{Ysard15-id, Nguyen18-ds, kalberla20-hm}: an increased dust-to-gas ratio associated with cold dense gas, which raises the FIR emission relative to $N_{\mathrm{H\,I}}$; optically thick \ion{H}{1} emission that lowers $N_{\mathrm{H\,I}}$ without affecting the associated dust emission; or spatially correlated molecular hydrogen ($\mathrm{H_2}$), which depletes the \ion{H}{1} population relative to the total hydrogen column. The caveat is that the different contributing effects are difficult to disentangle, but nevertheless they are all attributed to the cold phases of the ISM. \citet{Murray2020-uf} examined FIR/$N_{\mathrm{H\,I}}$ as binned by CNN-predicted $f_{\mathrm{CNM}}$ values, and found an enhancement of the ratio in higher-$f_{\mathrm{CNM}}$ bins. Similarly, here we examine the behavior of this ratio when binned by ST morphology measures. 

Following the procedures outlined in \citet{clark19_pn}, we apply a monopole correction of 0.64 $\mathrm{MJy\,sr^{-1}}$ \citep{planck16-mn}, from the Planck $I_{857}$ map, before projecting it onto the high-Galactic latitude ($\mathrm{b}>|30\degree|$) GALFA-\ion{H}{1} sky. In Figure \ref{fig:fn_hist}, we show the resulting histograms of $I_{857}/N_{\mathrm{H\,I}}^{90}$ in ST coefficient bins with equal numbers of sightlines. The ST coefficients are computed from $64\arcmin\times64\arcmin$ patches constructed around each pixel of the GALFA-\ion{H}{1} map, with a channel width $\delta v= 3$ km/s, and translated to per-sightline integrated measures using Equation \ref{eq:nhi_weight_st}, where $N_{\mathrm{H\,I}}(v)$ is the column density at a channel $v$ for a given pixel. Two such small-scale ST coefficients $S_\parallel\equiv S_{2\parallel}^{\mathrm{iso}}(j_1=0, j_2=1)$ and $S_\perp\equiv S_{2\perp}^{\mathrm{iso}}(j_1=0, j_2=1)$ are considered in the plot, converted to log scale and normalized to the range [0, 1]. From the discussion in Section \ref{subsec:st_reduc}, the $S_{\parallel}$ and $S_{\perp}$ coefficients can be interpreted as features aligning in parallel into sharp linear structures vs. features aligning perpendicularly into softer structures. The discussion in Section \ref{subsec:apply_st_intp} also shows that $S_\parallel$ correlates with $f_{\mathrm{CNM}}$, while $S_\perp$ shows strong anticorrelation. This is corroborated by the $I_{857}/N_{\mathrm{H\,I}}^{90}$ histograms in Figure \ref{fig:fn_hist}, where we see an enhancement of the ratio in larger bins of $S_\parallel$, while for $S_\perp$ the enhancement is in the opposite direction of smaller value bins, providing further evidence that small-scale linear structures are preferentially associated with cold phases of the ISM, while soft/diffuse structures are less likely to be found in regions of higher cold gas content. 

In Figure \ref{fig:fn_hist_regimes}, we compare $I_{857}/N_{\mathrm{H\,I}}^{90}$ binned by $N_{\mathrm{H\,I}}$, $S_\parallel/S_\perp$, and the $f_{\mathrm{CNM}}$ predicted by the CNN model in \citet{Murray2020-uf}, respectively, in two column density regimes: $N_{\mathrm{H\,I}}^{90}<4\times 10^{20}\,\mathrm{cm}^{-2}$ and $4\times 10^{20}\,\mathrm{cm}^{-2} < N_{\mathrm{H\,I}}^{90} < 1\times 10^{22}\,\mathrm{cm}^{-2}$. The degree of $I_{857}/N_{\mathrm{H\,I}}^{90}$ enhancement binned by the ST coefficients is comparable to that when binned with CNN-based $f_{\mathrm{CNM}}$ values in both regimes. The CNN-based $f_{\mathrm{CNM}}$ result is found to be consistent with \citet{Murray2020-uf}, when binned over the same column density range. In contrast, there is little enhancement with $N_{\mathrm{H\,I}}$ except in the case of very high column density, $N_{\mathrm{H\,I}}^{90}>10^{21}\,\mathrm{cm}^{-2}$. This behavior is consistent with results showing a fairly constant dust emission-$N_{\mathrm{H\,I}}$ ratio in the low column density regime $N_{\mathrm{H\,I}}<4\times 10^{20}\,\mathrm{cm}^{-2}$ \citep{lenz17-gd}. In this regime, the dust emission-to-\ion{H}{1} column ratio is consistent across the \ion{H}{1} column density values. At higher column densities, the increasing presence of the dust associated with $\mathrm{H_2}$ and the \ion{H}{1} column being less than the total hydrogen column, means that an extrapolation of the low column density linear correlation between $N_{\mathrm{H\,I}}$ and FIR underestimates the dust emission. This leads to enhancements of the $I_{857}/N_{\mathrm{H\,I}}$ ratio with $N_{\mathrm{H\,I}}$ at higher \ion{H}{1} column densities. Thus, in the low column density and linear dust emission-to-\ion{H}{1} column correspondence regime, the enhancement of the FIR/$N_{\mathrm{H\,I}}$ ratio with the small-scale linearity ST coefficient is a strong indication that the ST coefficient is predictive of CNM content. It should be noted that only a single small-scale ST coefficient is examined for the $I_{857}/N_{\mathrm{H\,I}}^{90}$ ratio study, so it is not representative of the full potential of the $f_{\mathrm{CNM}}$-correlating information that can be extracted from the morphology of the \ion{H}{1} emission. The multiple-correlation results in the previous sections, using $f_{\mathrm{CNM}}$ data from absorption measurements, suggest that a prediction using the full set of ST coefficients would produce further improved performance. 

\section{Discussion} \label{sec:discussion}
In this work, we explore the first application of the ST to \ion{H}{1} emission data. We demonstrate the utility of the ST for characterizing the \ion{H}{1} morphology, and connect our results to the important problem of \ion{H}{1} phase separation. 
Previous studies with GALFA-\ion{H}{1} data have identified highly linear filamentary structures in narrow \ion{H}{1} channel maps that are aligned with the plane-of-sky magnetic field orientation traced by dust and starlight polarization \citep{clark14-ma, clark15-fm}. These magnetically aligned \ion{H}{1} filaments are preferentially small-scale structures that are associated with the CNM \citep{clark14-ma, clark19_pn, peak19-cl, kalberla20-hm}. These previous results suggest that the spatial information in \ion{H}{1} maps potentially contains significant CNM-correlating information. In this work, rather than restricting our analysis to particular morphologies, like filamentary structures, we use the ST to explore a broader set of quantitative morphological descriptors. We present strong evidence of a correlation between \ion{H}{1} channel map emission morphology and $f_{\mathrm{CNM}}$. Our results are fully consistent with the picture of small-scale linear intensity structures being preferentially CNM, but the iterative application of the scattering operation allows us to probe complex scale and orientation interactions that describe more general morphological features in the \ion{H}{1} channel map data. Linearity is only one example of phase-correlating morphology, as suggested by the scale-dependent correlation behaviors discussed in Section \ref{subsec:apply_st_intp}.

Existing phase decomposition methods, be they traditional methods, like Gaussian decomposition \citep{matthews57-on, takakubo66-nh, mebold72-ot, haud07-gd, kalberla18-po, marchal19-rs, riener20-ag}, or more recent novel approaches, using CNNs \citep{Murray2020-uf}, mainly make use of spectral information to derive the phase content of the ISM. Some decomposition methods make use of neighboring pixel data, but only to ensure spatial coherence, not as a morphological indicator of phase \citep{marchal19-rs, taank22-mt}. Our model-independent correlation studies using the $f_{\mathrm{CNM}}$ data from absorption measurements suggest that the $f_{\mathrm{CNM}}$-correlating information content that can be extracted from spatial morphology may be comparable to the information that is contained in the spectral line shapes used by the CNN model prediction in \citet{Murray2020-uf}.

Our results thus suggest a way of improving phase decomposition methods, by making use of both the spatial and spectral structure of the \ion{H}{1} emission. In this regard, the problem of \ion{H}{1} phase decomposition may find synergies with problems like component separation of the cosmic microwave background (CMB) or data from line intensity mapping experiments. 
For example, GNILC (Generalized Needlet Internal Linear Combination) is a wavelet-based component separation algorithm that exploits both spectral information, in the form of the spectral energy density, and spatial information, in the form of angular power spectra \citep{Remazeilles11-gn, olivari16-gn}. The inclusion of spatial information, in particular, allows GNILC to distinguish emission components that suffer from spectral degeneracies, such as the cosmic infrared background and Galactic dust \citep{placnk20-gn}. Therefore, in order to achieve our goal of \ion{H}{1} phase separation, an important next step is to examine the orthogonality between spectral and spatial CNM-correlating information. If the morphology measures are significantly orthogonal to the spectral line shape, a combined approach would represent a clear improvement over current spectral decomposition methods. 

A combined spectral and spatial approach seems likely to be fruitful for constructing 3D (position-position-velocity) $f_{\mathrm{CNM}}$ maps. In particular, for constructing higher spectral resolution $f_{\mathrm{CNM}}$ maps, linewidth-based decompositions are limited by the spectral resolutions of the \ion{H}{1} emission data, with lower signal-to-noise ratios in narrower velocity channels. However, the narrow linewidths of CNM structures imply that their morphologies can be effectively quantified in narrow-channel maps \citep{clark14-ma}, consistent with our per-channel $f_{\mathrm{CNM}}(v)$ correlation results in Section \ref{subsec:corr_fcnm_v}. Thus, morphology measures of \ion{H}{1} emissions in narrow channels can complement spectral line information. Similarly, spectral information can be complementary to spatial information derived from morphological approaches using data with limited spatial resolutions. 

The existing simulations \citep[e.g.][]{Kim2014-db, kim17-ti} that were used to train the \citet{Murray2020-uf} CNN-based phase decomposition algorithm have high spectral resolutions, but limited spatial resolutions. Our work highlights the importance of having highly spatially resolved synthetic observations of the \ion{H}{1} 21 cm line from simulations. In addition to future applications building models capable of making accurate high resolution 3D $f_{\mathrm{CNM}}$ maps, such numerical simulations enable detailed comparisons between our work and theory. The morphologies of CNM structures contain imprints of a complex interplay between thermal instability, shock compression, turbulence, and magnetic fields at different scales. The connections between CNM morphology and the conditions of its multiscale turbulent ISM environment have been studied with numerical simulations \citep{hennebelle07-st, saury14-mh, inoue16-mh, gazol21-pg, fielding22-pm}. The ST, with its set of interpretable coefficients at different scales and orientations, can enable convenient quantitative comparisons between observations and simulations.
In different simulation contexts, well-motivated reductions to the ST coefficients can be adapted for different problems. For instance, ST linearity measures could be defined for specific orientations, like the magnetic field or shock propagation direction, to study how different conditions affect the formation and alignment of filamentary features.  

In addition to facilitating comparisons with numerical simulations, our results can be used for multiwavelength studies of ISM physics. Recent work utilizing the  \citet{Murray2020-uf} CNN-based approach to deriving an $f_{\mathrm{CNM}}$ map has demonstrated the connections between dust properties and CNM content, specifically a correlation between the fraction of dust in polycyclic aromatic hydrocarbons (PAHs) and $f_{\mathrm{CNM}}$ \citep{hensley22-pa}. The authors interpret these results as evidence that PAHs preferentially exist in cold and dense gas, likely because they are more easily destroyed in more diffuse gas. The \citet{hensley22-pa} paper also presented a new CNN-based $f_{\mathrm{CNM}}$ map from full-sky \ion{H}{1}4PI data that could be compared to our results, although the limited spatial resolution of \ion{H}{1}4PI (16.2\arcmin) compared to GALFA-\ion{H}{1} (4\arcmin) would restrict ST-based morphology characterization to larger angular scales than considered in this work.  

Finally, the technique described in this paper can be readily extended to study the \ion{H}{1} phase structure in different environments \citep[e.g.,][]{Murray:2021, Dickey:2022}. In addition to different Galactic environments, the recently released GASKAP-\ion{H}{1} Pilot Survey \citep{dempsey22-gs} of neutral hydrogen absorption in the Small Magellanic Cloud (SMC) provides a good testing ground for CNM-correlating ST morphology measures in a nearby dwarf galaxy. The physical scales probed by the GASKAP-\ion{H}{1} emission observations of the SMC are much larger than the physical extents of the small-scale Galactic CNM structures characterized here. 
It would be an interesting comparison to search for ST morphological signatures that are associated with the CNM at these larger scales. The ST can similarly be applied to \ion{H}{1} emission data from the Galactic plane, where the orientations of filamentary structures have been suggested to trace supernova feedback in the inner Galaxy, and Galactic rotation and shear in the outer Milky Way, using data from The \ion{H}{1}/OH/Recombination line survey \citep{soler20-gd, soler22-hm}. 

\section{Conclusions} \label{sec:conclusion}
In this work, we have shown that \ion{H}{1} emission morphology features extracted using the ST are highly predictive of the CNM fraction in the high-Galactic latitude ISM. Our main results are summarized as follows.
\begin{itemize}
    \item The ST is a powerful technique that can characterize the scale- and orientation-dependent information of fields in an efficient and interpretable way. We find that the ST is well suited to the task of deriving a set of morphological measures from \ion{H}{1} emission that are predictive of ISM properties, like the CNM content.
    \item We explore interpretable reductions of the ST coefficients, such as $S_\parallel$ and $S_\perp$, which describe features aligning along the same orientation into linear structures vs. those aligning along orthogonal orientations into diffuse structures. 
    \item We apply the ST to GALFA-\ion{H}{1} emission data and find that the small-scale $S_\parallel$ and $S_\perp$ coefficients are strongly correlated and anticorrelated with $f_{\mathrm{CNM}}$ data, respectively. These correlation trends, together with the interpretation of these coefficients as probing aligned parallel structures versus antialigned perpendicular structures, are consistent with the picture of high CNM content being associated with small-scale filamentary features, while regions with more diffuse and isotropic morphology have preferentially lower CNM fractions. 
    \item Model-independent correlation studies of the full sets of ST coefficients, with both LOS-integrated $f_{\mathrm{CNM}}$, and per-channel $f_{\mathrm{CNM}}(v)$ data, computed from absorption measurements, show that the ST coefficients contain significant CNM-correlating information. The degree of correlation is potentially comparable with the spectral information used by the CNN model in \citet{Murray2020-uf}. Both the spectral CNN method and the spatial ST method are more predictive of $f_{\mathrm{CNM}}$ than the column density alone.
    \item The link between the \ion{H}{1} emission morphology and the CNM mass fraction is further corroborated by the enhancement of the $I_{857}/N_{\mathrm{H\,I}}$ ratio, when binned by these small-scale ST coefficients. Regions with higher $S_\parallel$ and lower $S_\perp$ have higher $I_{857}/N_{\mathrm{H\,I}}$ than their surroundings, suggesting that they are associated with colder phases of the ISM.
    \item These results suggest that the ideal phase decomposition method would make use of both \ion{H}{1} emission spectral and spatial information to construct accurate 3D $f_{\mathrm{CNM}}$ maps. Toward that end, high-spatial-resolution multiphase numerical simulations and synthetic HI observations are needed to develop and test these decomposition models. 
\end{itemize}



\section{Acknowledgments}
We thank the anonymous referee for constructive feedback that helped improve the paper. We thank Claire E. Murray for helpful discussions. This publication utilizes the Galactic ALFA HI (GALFA-\ion{H}{1}) survey data set obtained with the Arecibo $L$-band Feed Array (ALFA) on the Arecibo 305 m telescope. The Arecibo Observatory is operated by SRI International under a cooperative agreement with the National Science Foundation (AST-1100968), and in alliance with Ana G. M\'endez-Universidad Metropolitana and the Universities Space Research Association. The GALFA-\ion{H}{1} surveys have been funded by the NSF through grants to Columbia University, the University of Wisconsin, and the University of California. This paper also makes use of observations obtained with Planck, an ESA science mission, with instruments and contributions directly funded by ESA Member States, NASA, and Canada. 

%

\vspace{5mm}


\software{Astropy \citep{astropy13},  
          scattering \citep{cheng2021-si}, 
          NumPy \citep{numpy11},
          scipy \citep{scipy20},
          matplotlib \citep{matplotlib07},
          PyTorch \citep{pytorch19}
          }



\appendix

\begin{figure}[t]
\centering
\includegraphics[width=\linewidth]{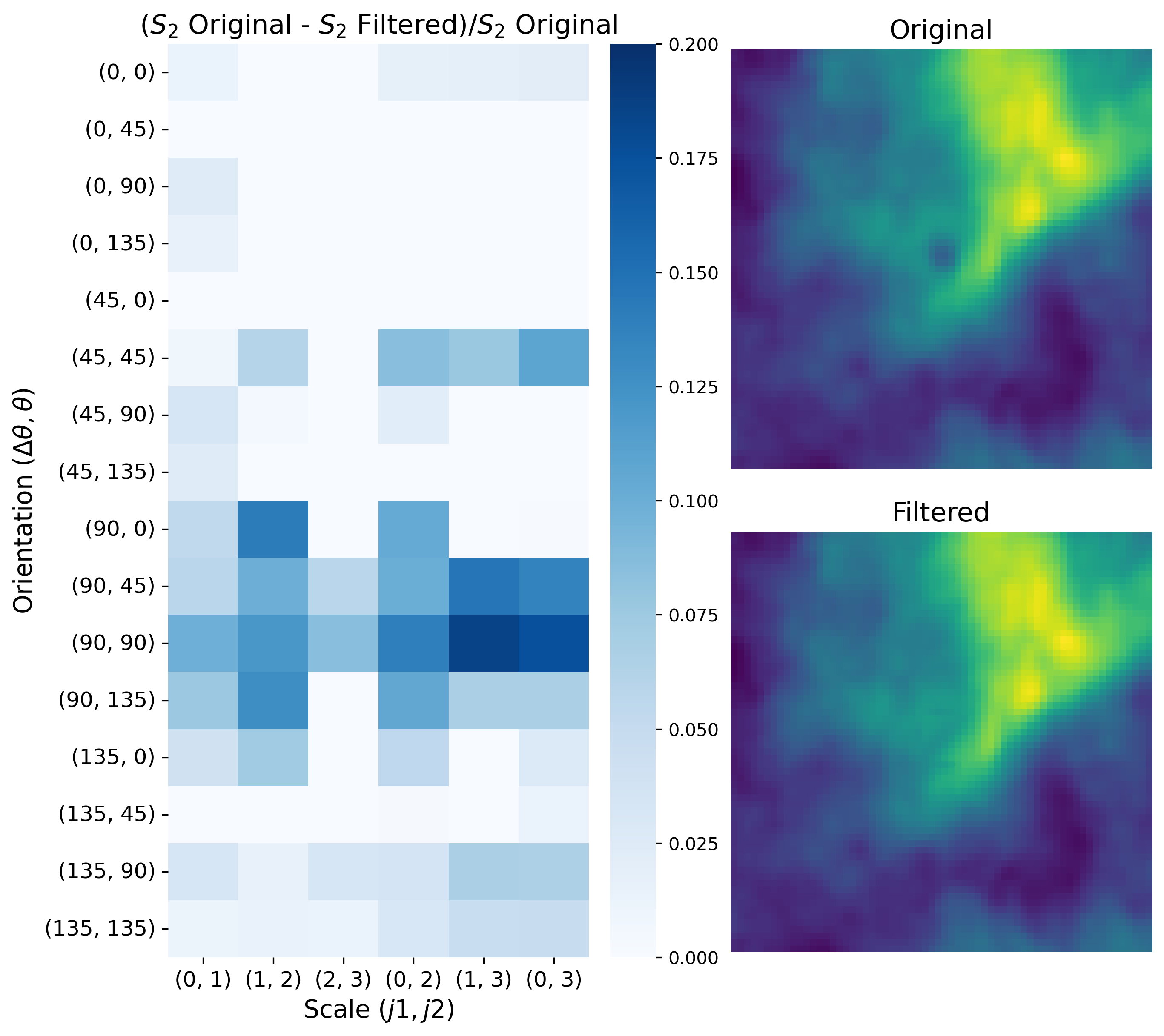}
\caption{Differences in the ST coefficients between a GALFA-\ion{H}{1} patch with a background continuum source and a version of the image after interpolating over the source. The original and filtered images are shown on the right. The left panel shows the relative differences between the ST coefficients for the original and filtered images, with the scale dimension along the x-axis and the orientation dimension along the y-axis. The original patch deviates the most from the filtered patch in its larger $S_{\perp}(\Delta\theta=90\degree)$ components aligning with the morphological interpretation that it is less linear overall, due to the central background source pixels.}
\label{fig:source}
\end{figure}

\begin{figure}[t]
\centering
\includegraphics[width=\linewidth]{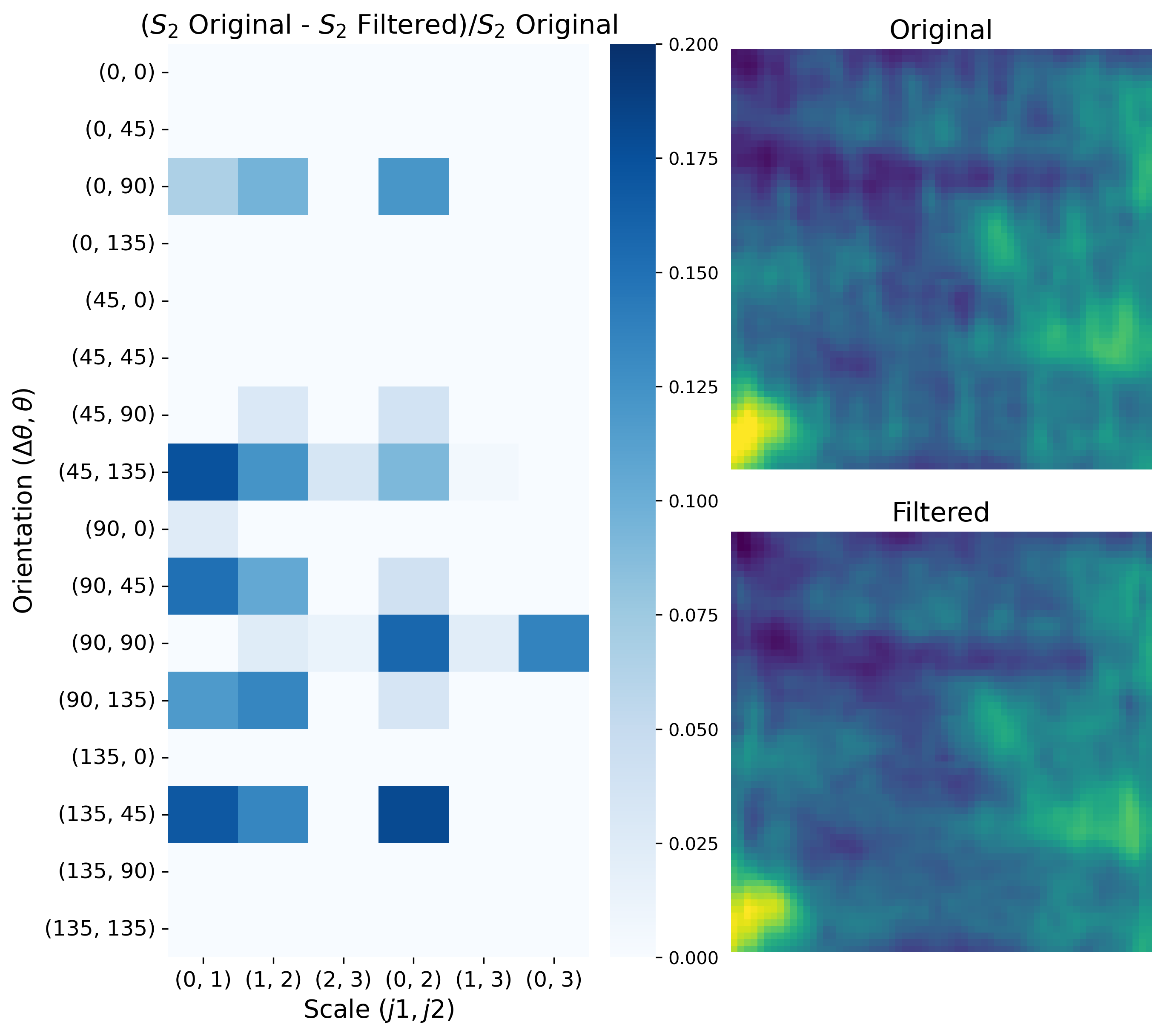}
\caption{Differences in ST coefficients between a GALFA-\ion{H}{1} patch with prominent scan pattern artifacts and a Fourier-filtered version. The original and filtered images are shown on the right. In the left panel, the relative differences between their ST coefficients are presented, with the scale dimension along the x-axis and the orientation dimension along the y-axis. The original patch deviates the most from the filtered patch in its larger $\theta=45\degree$ and $135\degree$ components aligning with the crisscross patterns of the scan artifacts.}
\label{fig:baseline}
\end{figure}

\begin{figure}[t]
\centering
\includegraphics[width=0.95\linewidth]{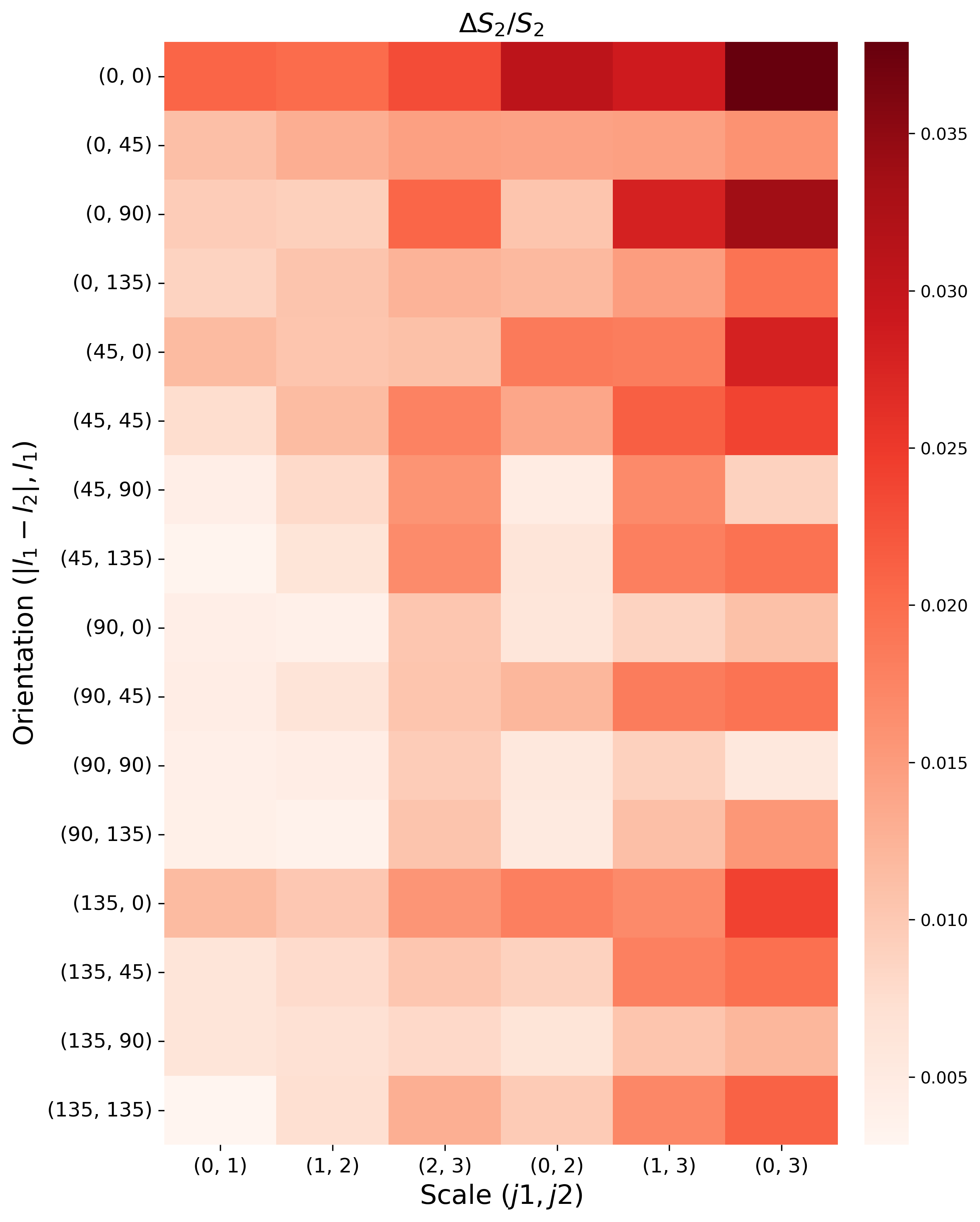}
\caption{The effect of adding ``pure noise" components to the GALFA-\ion{H}{1} patches on ST coefficient values. The resulting relative uncertainty $\Delta S_2/S_2$ averaged over all the patches, is shown for all scale and orientation components, as presented in Figures \ref{fig:st_val_corr} and \ref{fig:st_val_fcnm}. The deviations are within $\sim 4\%$ across the components, demonstrating that the results are not strongly sensitive to residual noise or systematics in the GALFA-\ion{H}{1} data.}
\label{fig:uq_value}
\end{figure}

\section{Effect of Systematics}\label{appx:systematics}
In this section, the effects of the GALFA-\ion{H}{1} data systematics on the ST coefficients are discussed in more detail. There are two main types of artifacts. First, since the GALFA-\ion{H}{1} patches used in our studies are constructed around sightlines with absorption measurements, the central pixels are occupied by background continuum sources, which appear as black dots on the images, with significantly lower intensity than the surrounding pixels. Secondly, artifacts that are associated with the basketweave scan pattern can manifest as image-space striping at known fixed angles \citep{Peek2017-ql}. As mentioned in Section \ref{subsec:apply_st_data}, we interpolate over the background source and Fourier filter to remove the fixed-angle rippling pattern. In Figures \ref{fig:source} and \ref{fig:baseline}, the effects of these systematics on ST coefficients are shown for two sample patches. We select the patches from the GALFA-\ion{H}{1} patches constructed around the absorption measurement sightlines, with a channel width $\delta v \sim 3$ km/s as used in the main analysis. We then apply the ST to the original and filtered versions of these patches, respectively, and present the relative differences in the resulting coefficients in a scale-orientation matrix, as first introduced in Section \ref{sec:hi_morph_res}. In both cases, the differences in the coefficient values between the original and filtered patches are within $\sim20\%$, and the components with the largest relative differences align with the expectations for the morphologies of the artifacts. In the background source case (Figure \ref{fig:source}), the original patch has smaller perpendicular coefficients $S_{\perp}(\Delta\theta=90\degree)$, due to the circular source pattern at the center of the patch. For patches with prominent telescope scan artifacts, the largest differences come from components with $\theta=45\degree$ and $135\degree$, corresponding to the crisscross ripple patterns.

To quantify the effects of the GALFA-\ion{H}{1} noise and artifacts on the $f_{\mathrm{CNM}}$ correlation studies, in Section \ref{sec:fcnm_corr_res} we add ``pure noise" components constructed from GALFA-\ion{H}{1} data in a high-velocity largely emission-free channel, centered at $v=400$ km/s with channel width $\delta v\sim3$ km/s. We estimate the resulting uncertainties using a Monte Carlo error propagation procedure. Note that this estimate of the uncertainty due to noise and artifacts should be conservative, since we apply pre-processing steps in the main analysis, including Fourier filtering, before calculating the ST coefficients. In Figure \ref{fig:uq_value}, we show the relative deviations of the $S_2$ coefficients averaged over the patches as a result of these noise components, in their full scale-orientation matrix form. These estimated uncertainties are incorporated into the correlation plots presented in Section \ref{sec:fcnm_corr_res}. From the plot, the deviations of the $S_2$ coefficients are within $\sim 4\%$. The small resulting relative uncertainties, despite the ST being applied to a small patch size of $N=64$ and narrow-channel width $\delta v\sim 3$ km/s, well demonstrate that our ST results are not strongly sensitive to residual noise or systematics in the GALFA-\ion{H}{1} data.

\section{Synthesized images with ST} \label{appx:st_synthesis}
\begin{figure*}[t]
    \centering
    \begin{minipage}{\linewidth}
        \centering
        \includegraphics[width=0.85\linewidth]{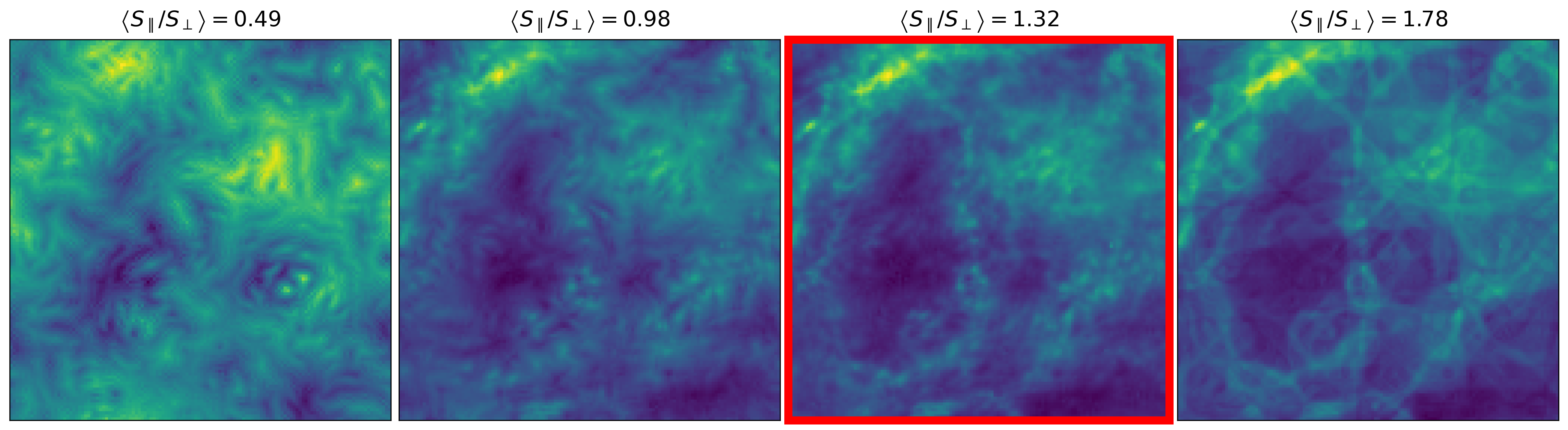}
        \caption{Synthesized patches with varying values of the ST coefficient $\left<S_{\parallel}/S_{\perp}\right>$. The patches are seeded from a sample GALFA-\ion{H}{1} patch that has the same ST values as the third image in the sequence. The images are shown with progressively larger values of $S_{\parallel}/S_{\perp}$, with the texture of the images going from more soft/diffused to more hard/linear, in line with the morphological interpretations discussed in Section \ref{sec:hi_morph_res}.}
        \label{fig:syn_avg}
    \end{minipage}
    \begin{minipage}{\linewidth}
        \centering
        \includegraphics[width=0.85\linewidth]{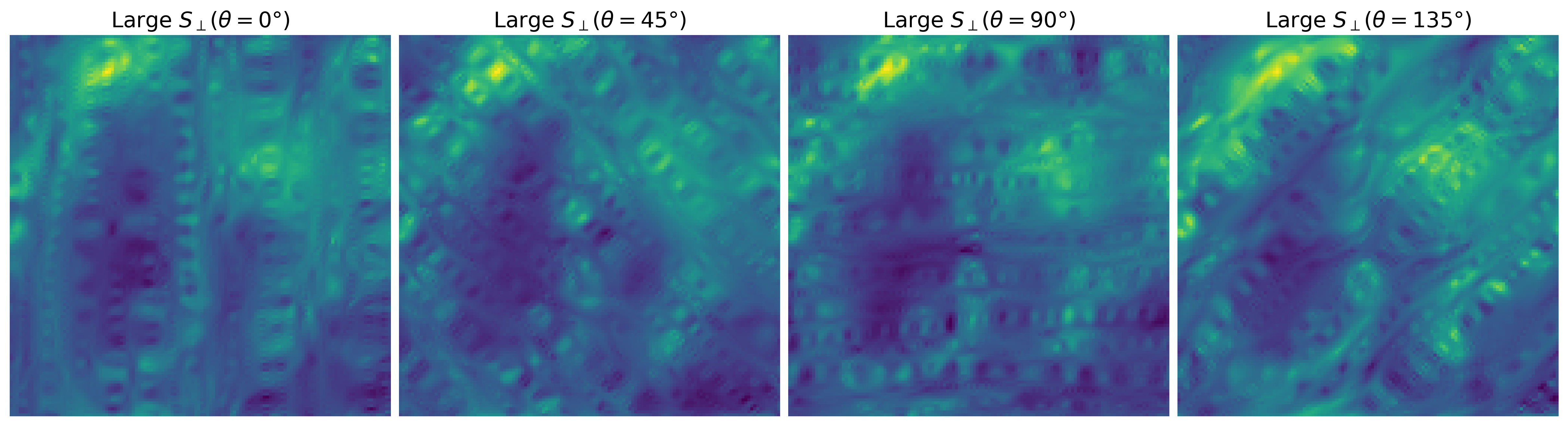}
        \caption{Synthesized images created from the same GALFA-\ion{H}{1} seed patch, while boosting a different orientation $\theta$ component of $S_{\perp}$. This results in the main textures aligning along different orientations for different images, consistent with the interpretation of the absolute orientation $\theta$. }
        \label{fig:syn_ori}
    \end{minipage}
    \begin{minipage}{\linewidth}
        \centering
        \includegraphics[width=0.87\linewidth]{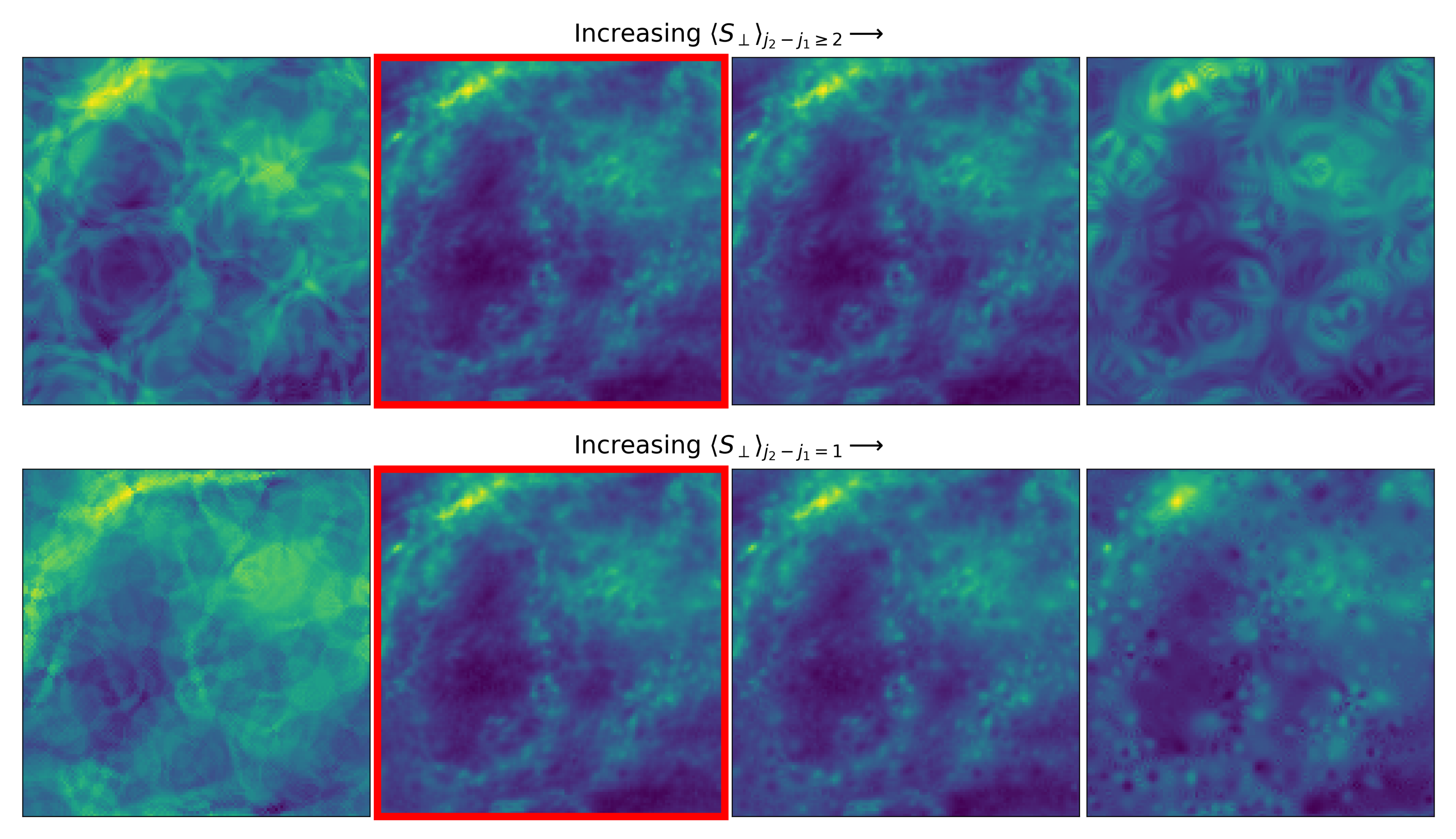}
        \caption{Synthesized images with varying values of $S_{\perp}$ at large scales $j_2-j_1\geq2$ (top) vs. small scales $j_2-j_1=1$ (bottom). The images are seeded from the same sample GALFA-\ion{H}{1} patch highlighted in red. The differences in texture between the corresponding images with varied small-scale vs. large-scale $S_{\perp}$ coefficients demonstrate the scale-dependent behaviors captured by the ST. For the rightmost images, with the largest $S_{\perp}$ values, the small-scale versions mainly consist of small-scale dots and clumps, while the $j_2-j_1\geq2$ version is populated by bending structures forming larger-scale circular features.}
        \label{fig:syn_scale}
    \end{minipage}
\end{figure*}

In Section \ref{subsec:st_reduc} we discussed a procedure for synthesizing images with a specific set of ST coefficients, showing representative patches in Figure \ref{fig:st_syn}. We implemented the minimization between the ST coefficients of the synthesized field with the target coefficients by using the ‘Adam’ optimizer in the python package $\texttt{torch.optim}$. Here, we present more such synthesized images to complement the discussion of the interpretations of the morphologies captured by different combinations of ST coefficients. In Figure \ref{fig:syn_avg}, we show synthesized patches with progressively larger values of $S_{\parallel}/S_{\perp}$, with the texture of the patches going from more soft/diffused to more hard/linear, in line with our interpretation for these coefficients in Section \ref{sec:hi_morph_res}. Then, in Figure \ref{fig:syn_ori}, we explore the interpretation of the absolute orientation component $\theta$. Each synthesized patch starts from the same seed image, as in Figure \ref{fig:st_syn}. A different $\theta$ component is then boosted for each patch, resulting in different images, where the main texture flows in different orientations. Finally, in Figure \ref{fig:syn_scale}, we explore the scale-dependent interpretation of the $S_{\perp}$ coefficients, motivated by the interesting scale-dependent $f_{\mathrm{CNM}}$-correlating behavior in Figure \ref{fig:st_val_fcnm}, where $S_{\perp}$ anticorrelates with $f_{\mathrm{CNM}}$ at small scales, but positively correlates at large scales. On the top panels, four synthesized patches are shown, with increasing values of $S_{\perp}$ on scales $j_2-j_1\geq 2$, contrasted with the lower panels, which show corresponding patches at increasing values of $S_{\perp}$ on small scales $j_2-j_1 = 1$. Both cases are synthesized from varying the ST coefficients of the same seed image. As $S_{\perp}$ values are varied along either direction for the other images, clear qualitative differences in texture can be seen between the small- and large-scale versions. In particular, for the images on the right with the largest $S_{\perp}$ values, the small-scale case mainly consists of small-scale dots and clumps, while the $j_2-j_1\geq2$ version is populated by bending structures forming larger-scale circular features. This complexity of features demonstrates the ability of the ST to probe and capture complex orientation- and scale-dependent behaviors. 


\onecolumngrid 

\clearpage
\bibliography{st4cnm}{}
\bibliographystyle{aasjournal}



\end{document}